# Model-assisted estimation for MRV

How to boost the economics of SOC sequestration projects without compromising on scientific integrity

**Ahmad Awad[1] and Erik Scharwächter[1]**

[1] Seqana GmbH, Berlin, Germany

**Correspondence**: Erik Scharwächter (erik.s@seqana.com)

## Executive Summary

- Monitoring, Reporting, and Verification (MRV) in SOC sequestration projects requires statistical estimation approaches that follow the **highest standards of scientific integrity** to avoid overcrediting and build trust in carbon markets.

- Model-assisted estimation is a cost-effective way to **increase the precision of SOC stock estimates** while preserving scientific integrity. Increased precision leads to smaller sample sizes and lower uncertainty deductions, which substantially improves project economics.

- In contrast to model-based estimation (e.g., Verra's VT0014), model-assisted estimation **does not require project-specific model validation** to prove scientific integrity. Our simulation study demonstrates that the simple regression estimator (SRE, a model-assisted estimator) yields unbiased SOC stock estimates and valid uncertainty estimates once sample sizes exceed 40 samples, across a wide variety of hypothetical SOC sequestration projects with different levels of SOC stock variability.

- The precision gains obtained by SRE are **directly proportional to the predictive power of the model**. Theoretical and empirical results show that regression models with $r^2 = 0.3$ reduce uncertainties by 30%. If regression models lack predictive power, SRE performs comparable to standard design-based methods, when sample sizes exceed 40 samples.

- **Recommendations**: Regulators should update methodologies to explicitly allow model-assisted estimators, with effective minimum sample size requirements. Project developers should routinely employ SRE or other model-assisted estimators if models/covariates with high predictive power are available for their project areas. Industry stakeholders should prioritize investment in strong models/covariates to maximize economic efficiency of SOC sequestration projects.

# Contents

# Abstract

Soil organic carbon (SOC) sequestration projects require unbiased, precise and cost-effective Monitoring, Reporting, and Verification (MRV) systems that balance sampling costs against uncertainty deductions imposed by regulatory frameworks. Design-based estimators guarantee unbiasedness but cannot exploit auxiliary data. Model-based approaches ([VCS_Methodology_VT0014_v1.0 (2025)](#)) can improve precision but require independent validation for each project. Model-assisted estimation offers a robust compromise, combining model predictions with probability sampling to retain design-based guarantees while improving precision.

We evaluate the scientific integrity and efficiency of the simple regression estimator (SRE), a well-known model-assisted estimator, via an extensive simulation study. Our simulations span diverse SOC stock variances, sample sizes, and model performances. We assess three core properties: empirical bias, empirical confidence interval coverage, and precision gain relative to the design-based Horvitz-Thompson estimator (HTE). Results show negligible bias and valid coverage probabilities for n > 40, regardless of SOC stock variance. Below this threshold, variance approximations and normality assumptions yield unreliable uncertainty estimates. With correlated ancillary variables ($r^2$ = 0.3), SRE achieves 30% precision gains over HTE. With uncorrelated variables, no gains are observed, but performance converges to HTE for n ≥ 40.

Model-assisted estimation can enhance project economics without compromising scientific rigor. Regulators should permit such estimators while mandating minimum sample size thresholds. Project proponents should routinely employ such estimators when correlated ancillary variables exist. The industry should prioritize the retrieval of high-quality, project-specific covariates to maximize precision gains and thereby the project economics.

# 1. Introduction

The **economic viability** of SOC sequestration projects is directly impacted by the number of soil samples that must be taken on the ground for effective Monitoring, Reporting and Verification (MRV). On the one hand, sample sizes determine costs of the soil sampling campaign(s). On the other hand, they strongly affect the uncertainty deductions to be applied on the SOC stock change estimates. Uncertainty deductions are often overlooked during project scoping, but remain an integral part of all methodologies governing the voluntary carbon market (VCM) and Scope 3 MRV. They are ultimately determined by the precision of SOC stock (change) estimates calculated for the project. For this reason, **improving the precision** of these estimates is a key lever to boost project economics. In the past years and decades, considerable efforts have been made in both industry and academia to improve the precision of SOC stock (change) estimates by exploring combinations of remote sensing, machine learning, process-based modeling, and optimized sampling designs for ground-measured data. Many of these efforts have demonstrated promising results. However, the specific technical details, underlying assumptions, and data requirements associated with these methods often present substantial challenges for their direct application to SOC sequestration projects operating within regulated carbon markets that require provable **scientific integrity**.

For scientifically rigorous SOC MRV, two requirements are critical to control the risks of generating credits for carbon that was never truly sequestered. These requirements address overcrediting due to **systematic estimation errors**, and overcrediting due to **random estimation errors**, respectively:

1. The **SOC stock (change) estimates** calculated for a project must be **unbiased**. Unbiased estimators avoid systematic overestimation (and underestimation) of SOC stock change that can lead to systematic overcrediting (or undercrediting) in the project.
2. The **uncertainties** calculated for the SOC stock (change) estimates must be **statistically valid**. This means they must reliably capture the random variation of the reported figures. Reliable uncertainty estimates enable the methodological safeguard of uncertainty deductions to effectively reduce the risk of overcrediting due to random variation in the estimates.[1]

---

[1] Strictly speaking, only the effect size estimates, i.e., estimates for the differences between project SOC stock change and baseline SOC stock change need to be unbiased and come with statistically

Novel MRV technologies that aim at improving the precision of SOC stock (change) estimates must be supported by sufficient evidence demonstrating that these two requirements for scientific integrity are met. For example, VCS Methodology VM0042 v2.1 (2024) allows SOC stock change estimation using process-based models, provided that the two requirements above are empirically validated on data that is representative of the project context (c.f. VCS Methodology VMD0053 v2.1 (2025)). Similarly, geostatistical models (digital soil maps) may be used for SOC stock change estimation, but must be validated on data taken from the project area at project start and all subsequent verification events (c.f. VCS Methodology VT0014 v1.0 (2025)). The reason why these **model-based estimation** approaches need to be validated empirically is that the scientific integrity of the resulting SOC stock change estimates is only guaranteed if the underlying modeling assumptions are met in the project context. Preparing a model validation report in line with the methodological requirements to confirm scientific integrity is laborious, costly and may not be economically feasible for some project proponents.

An underexplored alternative to model-based estimation is the direct integration of model predictions into **design-based estimation** approaches. Design-based inference provides estimates for SOC stock (change) from soil samples taken in compliance with a probability sampling design.[2] All of today's methodologies for SOC MRV prescribe design-based estimators at some point. The well-known "measure and re-measure" quantification approach based on soil sampling is, in fact, a purely design-based quantification approach, whereas "measure and model" uses design-based estimation to initialize a process-based SOC model for subsequent model-based estimation. Many design-based estimators are mathematically proven to be unbiased and come with procedures to obtain statistically valid uncertainty estimates (de Gruijter et al. 2006). The most well-known design-based estimator that satisfies these two properties is the sample mean, provided that samples are taken following a simple random sampling design. Most methodologies recommend the use of a stratified simple random sampling design and estimator, since it typically yields lower uncertainties than simple random sampling.[3]

---

valid uncertainty estimates. However, requiring unbiasedness and valid uncertainties already on the level of SOC stock and SOC stock change estimates contributes to increased scientific rigor.

[2] The word "design" in design-based estimation refers to the underlying sampling design.

[3] On a more fundamental level, the difference between design-based and model-based inference is the source of randomness that is considered in the statistical equations. Design-based inference takes a probability sampling design as the source of all randomness and views the SOC distribution on the ground as fixed. In contrast, model-based estimation approaches assume that the SOC distribution on the ground was produced by a random process and that the sampling locations are

The key advantage of design-based inference over model-based inference is that the properties of design-based estimators **hold in any SOC sequestration project** as long as the project proponent strictly follows the prescribed probability sampling design. In other words: The scientific integrity of a design-based estimator employed within a SOC sequestration project is guaranteed by the correct implementation of the soil sampling campaign, regardless of any other project context.

The integration of model predictions into a design-based estimation framework is referred to as **model-assisted estimation** (Cochran 1977; Särndal et al. 1992; de Gruijter et al. 2006; Breidt and Opsomer 2017). What makes model-assisted estimation attractive is that its scientific integrity is maintained regardless of whether the assumptions of the model are met. In fact, the scientific integrity of a model-assisted estimate still depends on the correct implementation of the sampling design. As long as project proponents adhere to the prescribed sampling design, it is not necessary to validate the model empirically before using it in the project context. In particular, current regulatory frameworks that allow model-assisted estimation (e.g., Gold Standard (2025), Australian Government (2025)) do not require prior validation of the utilized model. This makes model-assisted estimation a promising, cost-effective alternative to model-based estimation to increase precision of SOC stock (change) estimates.

There is a caveat: **asymptotics**. While the scientific integrity of design-based SOC stock (change) estimates is guaranteed at any sample size, model-assisted estimates only have asymptotic guarantees. This means that the estimates may be biased at small sample sizes, but become practically unbiased starting from a certain minimum sample size. The same holds true for the calculated uncertainties, which correctly capture the random variation of the estimates only if the sample size is not too small.[4] To strengthen the case

---

fixed. The random variation of a design-based estimator is measured over many hypothetical realizations of the sampling design, whereas random variation of a model-based estimator is measured over many hypothetical SOC distributions on the ground. In theory, both approaches only yield scientifically rigorous estimates if their underlying assumptions are met. However, it is much easier to take soil samples in compliance with a fully specified probability sampling design than to reverse engineer the random process that generated the SOC distribution in the project area. A thorough discussion of (geostatistical) model-based and design-based inference can be found in (Brus and de Gruijter 1997).

[4] Depending on the complexity of the utilized model, overfitting may exacerbate these issues. The effect of overfitting and several safeguards against it (for smaller sample sizes) will be discussed in detail in a follow-up paper.

for the adoption of model-assisted estimators within regulated carbon markets, it is essential to better understand the practical implications of these asymptotic guarantees. Can model-assisted estimators provide higher precision SOC stock (change) estimates without compromising on scientific integrity?

To address this question, this work empirically studies the asymptotic behavior of the **simple regression estimator (SRE)**, a well-known and widely used model-assisted estimator ([Cochran 1977](#); [Brus 2000; 2023](#)). In particular, the focus of our simulation study lies on finding a minimum sample size at which the scientific integrity of SRE can be compared to that of the **Horvitz-Thompson estimator (HTE)**, the default design-based estimator that does not make use of model predictions. We also look at the empirical precision gain achieved by SRE compared to HTE, for models with different explanatory powers.

## 2. Statistical Background

Model-assisted estimation is a well-established statistical method that can reduce SOC stock (change) estimation uncertainties in a design-based estimation framework. The key idea is to combine predictions from a so-called "working model" with ground-measured SOC stock values coming from a probability sampling design. The higher the explanatory power of the working model, the higher the precision gains due to the model-assisted estimator ([Särndal et al. 1992](#)). In the past decades, many model-assisted estimators have been derived, and general recipes exist to construct them for any working model and probability sampling design ([Breidt and Opsomer 2017](#)). A common property of all model-assisted estimators is that they can be decomposed into two components:

1. a prediction of the target variable (mean SOC stock) based on the working model,
2. a bias-correction term that captures systematic prediction errors of the working model and is estimated from the probability sample.

The working model determines which statistical relationships between the ancillary variables (covariates) and the target variable can be exploited for estimation (e.g., linear or nonlinear). The most well-studied model-assisted estimator suitable for any probability sampling design is the generalized regression estimator (GREG). It is based on a heteroscedastic multiple linear regression working model ([Breidt and Opsomer 2017](#); [Särndal et al. 1992](#)). GREG is known to be asymptotically design-unbiased and comes with an approximately valid uncertainty estimator ([Breidt and Opsomer 2017](#)). In this work, we study a special case of GREG that uses a homoscedastic simple linear regression working model and ground measurements taken with simple random sampling ([Cochran 1977](); [Brus 2000](); [2023]()). We refer to this estimator as the simple regression estimator (SRE). The notation and explanations largely follow ([Brus 2023]()) with some minor amendments.

## 2.1. Simple regression estimator

Let $z_1, ..., z_n$ denote a set of $n$ ground-measured SOC stock values taken with simple random sampling within a project area. The **Horvitz-Thompson estimator** (HTE, design-based) for the mean SOC stock in the project area is the sample mean $\bar{z}_S$ of the ground-measured values:

$$\hat{\bar{z}}_{HTE} = \bar{z}_S = \frac{1}{n} \sum_{k=1}^{n} z_k \tag{1}$$

This estimator is unbiased for any $n$. Let now $x_1, ..., x_n$ denote the corresponding values of some ancillary variable at the same $n$ sampling locations. The working model of SRE assumes that the target variable can be expressed as a simple linear function of the ancillary variable with (unknown) intercept α, slope β, and a zero-centered residual $\epsilon_k$:

$$Z_k = \alpha + \beta x_k + \epsilon_k \tag{2}$$

The slope and intercept parameters of the working model are estimated from the probability sample using ordinary least squares, which yields:

$$\hat{b} = \frac{\sum_{k=1}^{n}(x_k - \bar{x}_S)(z_k - \bar{z}_S)}{\sum_{k=1}^{n}(x_k - \bar{x}_S)^2} \tag{3}$$

$$\hat{a} = \bar{z}_S - \hat{b} \cdot \bar{x}_S \tag{4}$$

In these equations, $\bar{x}_S$ denotes the sample mean of the ancillary variable computed from the $n$ values of the ancillary variable at the sampled locations. Let now $\bar{x}$ denote the population mean of the ancillary variable within the project area.[5] The **simple regression estimator** (SRE) for the mean SOC stock in the project area is defined as:

$$\hat{\bar{z}}_{SRE} = \bar{z}_S + \hat{b} \cdot (\bar{x} - \bar{x}_S) \tag{5}$$

---

[5] Model-assisted estimation requires that the values of all ancillary variables are known across the complete project area. This is typically the case for remote sensing-based datasets (e.g., vegetation indices) as well as many other geospatial datasets (e.g., digital elevation models) that are distributed in a rasterized format. The population mean of an ancillary variable within the project area can easily be computed by averaging all raster cells that lie within the project area.

Equation 5 shows that SRE adjusts HTE by a term that depends on the estimated slope and the ancillary variable means. The estimator can be rewritten more verbosely as

$$\widehat{\bar{z}}_{SRE} = \hat{a} + \hat{b} \cdot \bar{x} + \frac{1}{n}\sum_{k=1}^{n} e_k \qquad (6)$$

to reveal that it decomposes into (1) a model prediction $\hat{a} + \hat{b} \cdot \bar{x}$ for the population mean and (2) a bias-correction term estimated from the residuals of the linear regression model $e_k = z_k - (\hat{a} + \hat{b} \cdot x_k)$. Since SRE is a special case of GREG, it is known to be asymptotically unbiased, but not unbiased for every $n$. The reason why it is only asymptotically unbiased is that the intercept and slope parameters are estimated from the same ground-measured data as the bias-correction term. The simulation study carried out below analyzes the practical impact of this property on the scientific integrity of SRE estimates.

## 2.2. Uncertainty of the simple regression estimator

The main reason to employ a model-assisted estimator like SRE is that it can reduce the uncertainty of a SOC stock estimate compared to HTE. Higher precision SOC stock estimates translate to higher precision SOC stock change estimates and higher precision effect size estimates (project change minus baseline change). This means that project proponents need fewer soil samples to obtain an estimate with the same precision, or that they can apply lower uncertainty deductions with the same sample size. In both cases, the SOC sequestration project becomes more profitable.

The uncertainty of an estimator is measured via the standard error, i.e., the square root of its sampling variance. The larger the sampling variance, the higher the uncertainty. The **sampling variance of HTE** (design-based) from Equation 1 can be estimated as

$$\widehat{V}(\widehat{\bar{z}}_{HTE}) = \frac{1}{n}\widehat{S^2}(z) \qquad (7)$$

where $\widehat{S^2}(z)$ is an estimate for the SOC stock variance based on the ground-measured SOC stock data. This estimate of the sampling variance is valid for any sample size. For comparison, the **sampling variance of SRE** (model-assisted) can be approximated by

$$\widehat{V}(\widehat{\bar{z}}_{SRE}) = \frac{1}{n}\widehat{S^2}(e) \qquad (8)$$

where $\widehat{S^2}(e)$ is an estimate for the variance of the regression residuals, based on the same ground-measured data. SRE thus yields lower uncertainties than HTE if the regression residuals have a lower variance than the raw ground-measured values ([Breidt and Opsomer 2017](#)). For the simple linear regression model, the variance of the regression residuals is directly proportional to the squared correlation $r^2$ between the ancillary variable and SOC stocks ([Särndal et al. 1992](#)). The sampling variance of SRE can thus be expressed approximately via the sampling variance of HTE:

$$\widehat{V}\left(\widehat{\bar{z}}_{SRE}\right) = (1 - r^2)\, \widehat{V}\left(\widehat{\bar{z}}_{HTE}\right) \qquad (9)$$

This result demonstrates that the reduction in uncertainty is larger the stronger the correlation between the ancillary variable and SOC stocks. However, the approximation from Equation 8 is only valid if the sample size is not too small. A better estimator for the sampling variance of SRE is the g-weight estimator ([Hill et al. 2021](#)), which takes the uncertainty of the regression parameters into account. The practical implications of these approximations on the scientific integrity of SRE are studied empirically below.

## 2.3. Other model-assisted estimators

SRE can easily be extended to a stratified simple random sampling design ([Cochran 1977](#); [Viscarra Rossel et al. 2016](#); [Brus 2023](#)). In this case, the working model can either be fitted separately for each stratum or jointly across all strata. This means that no deviations from established sampling protocols are required when adopting a model-assisted estimation approach. Combining stratified sampling with model-assisted estimation is particularly useful if the stratification variable alone does not adequately capture the SOC stock variability on the ground, and additional ancillary variables correlated with SOC stock can be employed for model-assisted estimation. As pointed out above, many more model-assisted estimators exist apart from the GREG family of estimators, including estimators based on linear mixed models, kernel methods, splines, neural networks, k-nearest neighbors, and others ([Breidt and Opsomer 2017](#)).

# 3. Methods

We perform a simulation study to verify the scientific integrity of SRE and its suitability for MRV in SOC sequestration projects. Concerns about the scientific integrity of model-assisted estimators may be raised on the basis of their asymptotic unbiasedness and the approximative nature of the sampling variance estimator, in particular, for smaller sample sizes. Therefore, key questions that we try to answer with our simulations are:

1. At what sample sizes can SRE be viewed as **practically unbiased**?

2. At what sample sizes can the approximate uncertainty estimates obtained for SRE be seen as **practically valid**?
3. What are the expected **precision gains** of SRE compared to HTE?

To capture a wide range of real-world conditions, we study these questions for simulated SOC sequestration projects with varying total population variances $Var(Z)$, reflecting different environmental and management scenarios. We systematically increase the sample size to be used for both design-based (HTE) and model-assisted estimation (SRE) of SOC stocks. At each sample size, we apply SRE twice—once using a correlated ancillary variable (SRE-corr) and once using an uncorrelated ancillary variable (SRE-uncorr)—to reflect differing levels of auxiliary information quality. We then identify the sample size at which SRE yields estimates of a high scientific integrity across all scenarios. We also study the precision gains that can be achieved across all scenarios.

### 3.1. Population model

We simulate SOC sequestration projects by sampling the spatial distribution of SOC stocks from a geostatistical model. Let $Z$ denote the SOC stocks in the hypothetical project area. We assume that $Z$ can be decomposed into a sum of two quantities,

$$Z = X + \Delta \qquad (10)$$

The quantity $X$ is assumed to be known throughout the complete project area and is taken as the correlated ancillary variable for SRE-corr. The residual $\Delta$ captures all SOC stock variation that cannot be explained by the correlated ancillary variable. We use spherical covariance models to generate both $X$ and $\Delta$. These covariance models are parameterized by (1) the sill, defined as $Var(X)$ and $Var(\Delta)$, respectively, (2) the spatial autocorrelation range, denoted as $r_X$ and $r_\Delta$, and (3) the nugget effect that we set to zero. The predictive power of $X$ for $Z$ can be controlled by the altering the ratio between the variance of $X$ ("explained variance") and the variance of $\Delta$ ("unexplained variance").

We additionally generate a quantity $X_{uncorrelated}$ that is completely uncorrelated with the target variable $Z$, by sampling from a normal distribution

$$X_{uncorrelated} \sim N(0, Var(X_{uncorrelated})) \qquad (11)$$

and decorrelating the sampling result from $Z$ by removing any spurious trends that are due to the random sampling process. The decorrelated quantity is used to study the performance of SRE with an uncorrelated ancillary variable (SRE-uncorr).

In practice, the variance of SOC stocks may range from 100 to 2100 tC²/ha². Five equal increments of 400 tC²/ha² were used to retrieve a total of **6 simulated SOC sequestration projects.** The ratio between explained and unexplained variance was controlled such that the squared correlation between $X$ and $Z$ was $r^2 \approx 0.3$.

Since we use simulated SOC sequestration projects instead of real projects for our analysis, we have access to the true SOC stock population means (not only estimates thereof). The true population mean is required to correctly assess the bias of an estimator, as well as the coverage probability of its confidence interval.

## 3.2. Sampling and estimation

For all 6 simulated SOC sequestration projects, we simulate soil sampling campaigns with an increasing number of samples. For each sample size, we fit SRE-corr using the ancillary variable $X$ and SRE-uncorr using the ancillary variable $X_{uncorrelated}$. We use both variants to obtain model-assisted estimates for the mean SOC stock over the project area. As a baseline for comparison, we additionally use HTE to obtain design-based estimates for the mean SOC stock. In the simulated sampling campaigns, we use a simple random sampling (SRS) design with sample sizes

$$n \in \{5, 10, 20, 40, 80, 160, 320, 640, 1280, 2560, 5120\}.$$

To obtain stable performance metrics, we perform $M = 10000$ Monte Carlo replicates of the sampling and estimation approach, for each SOC sequestration project and sample size. Overall, this setup yields a total of 198 scenarios (combinations of SOC sequestration project, sample size, and estimator) with 1,980,000 estimation results to evaluate.

In each Monte Carlo replicate $m$, we perform the following:

1. Draw a simple random sample (SRS) of size $n$.
2. Estimate $\hat{\bar{z}}^{(m)}$ with both SRE variants (Equation 5, Section 2.1) and with HTE (Equation 1, Section 2.1).

3. Estimate the sampling variance $\widehat{V}^{(m)}(\widehat{\bar{z}})$ for HTE using Equation 7, Section 2.2.
4. Estimate the sampling variances $\widehat{V}^{(m)}(\widehat{\bar{z}})$ for both SRE variants using the g-weight estimator (Hill et al. 2021).
5. Estimate confidence intervals $CI^{(m)}$ for all estimators, using the approach outlined below.

Confidence intervals for all estimators were obtained under a normality assumption by rescaling the standard errors of the estimators with appropriate t-values for the desired nominal coverage probability:

$$CI^{(m)} = \widehat{\bar{z}}^{(m)} \pm t_{1-\alpha/2} \sqrt{\widehat{V}^{(m)}(\widehat{\bar{z}})} \qquad (12)$$

where $\widehat{\bar{z}}^{(m)}$ is either $\widehat{\bar{z}}_{SRE}$ or $\widehat{\bar{z}}_{HTE}$ defined in Equation 1 or 5, respectively, applied on the $m$-th Monte Carlo replicate. We use $\alpha = 0.05$ to obtain confidence intervals with a nominal coverage probability of 95%.

### 3.3. Performance metrics

To evaluate the scientific integrity of SRE and compare it with HTE, we focus on two key metrics: empirical bias of the SOC stock mean estimates, and empirical coverage probability of the estimated confidence intervals. Moreover, we compare the precision of all estimates to better understand the precision gains that can be achieved with SRE.

Unbiasedness is paramount to retrieving trustworthy estimates that do not systematically under- or overestimate the mean SOC stock. To assess at what sample size SRE can be viewed as practically unbiased, we compute the **empirical bias** of the estimator based on $M$ Monte Carlo replicates as follows:

$$\widehat{Bias}_M = \frac{1}{M} \sum_{m=1}^{M} \left( \widehat{\bar{z}}^{(m)} - \mu \right) \qquad (13)$$

where $\mu$ is the (known) population mean of SOC stock in the simulated project area. The estimator can be viewed as practically unbiased if the empirical bias is close enough to 0.

To assess whether the empirical bias is close enough to 0, we perform two-tailed t-tests against the null hypothesis of zero bias.

The **empirical confidence interval coverage probability** allows evaluating the validity of the uncertainty estimates computed for an estimator. Since the sampling variance estimators for SRE are only approximative, it is important to understand at what sample sizes they yield reliable results. The empirical coverage probability based on $M$ Monte Carlo replicates is estimated as follows:

$$\widehat{C}_M = \frac{1}{M} \sum_{m=1}^{M} 1\{\mu \in CI^{(m)}\} \tag{14}$$

The uncertainty estimate is considered to be practically valid if the empirical coverage probability is at least as high as the nominal coverage probability of 95%.

At last, we study the **precision gain** of SRE compared to HTE to assess whether and how much it can reduce SOC stock estimation uncertainties. Precision gain (also referred to as design effect or stratification effect, depending on context) provides a measure of the efficiency of a probability sampling design and/or estimator relative to other estimators and/or probability sampling designs ([Lohr 2019](#); [Brus 2023](#)). Formally, we define the empirical precision gain $\widehat{Gain}_M$ as the ratio between the Monte Carlo sampling variances of SRE and HTE:

$$\widehat{Gain}_M = \frac{\widehat{V}_M(\widehat{\bar{z}}_{SRE})}{\widehat{V}_M(\widehat{\bar{z}}_{HTE})} \tag{15}$$

$$\widehat{V}_M(\widehat{\bar{z}}) = \frac{1}{M-1} \sum_{m=1}^{M} \left( \widehat{\bar{z}}^{(m)} - \frac{1}{M} \sum_{m'=1}^{M} \widehat{\bar{z}}^{(m')} \right)^2 \tag{16}$$

A precision gain less than 1 indicates that SRE is more precise than HTE, while a precision gain greater than 1 indicates the opposite. We use Monte Carlo sampling variances to calculate the precision gain to make sure the precision gain metric is not affected by inaccurate sampling variance estimators, especially at small sample sizes.

# 4. Results

## 4.1. Empirical bias

Overall, our simulation results demonstrate that SRE is practically unbiased even for the smallest sample sizes. The two-tailed t-test revealed that the empirical bias generally remains statistically insignificant (at a 5% significance level) across most combinations of SOC stock variances and sample sizes. This observation holds regardless of whether a correlated or an uncorrelated ancillary variable is used in the working model (see Table 1 in Appendix I). The t-test rejected the null hypothesis of unbiasedness only in a few cases where the SOC stock variance was high and sample sizes were very small ($n \leq 20$).

In total, 11 out of 198 t-tests (5.5%) performed across all scenarios rejected the null hypothesis of unbiasedness. This is close to the expected Type I error rate of 5% and may be due to random variation only. However, almost all statistically significant biases (8 out of 11) were detected for the scenario with the largest SOC stock variance (2,100 tC²/ha²). This indicates a systematic pattern that cannot be explained by random variation alone. 3 out of 66 tests (4.5%) rejected the null hypothesis of unbiasedness for both HTE and SRE-uncorr, while 5 out of 66 tests (7.6%) rejected for SRE-corr.

Figure 1 visualizes the bias evaluation results along with critical values for the t-test, at two different simulated SOC stock variances, for the three estimators and various sample sizes. At a SOC stock variance of 900 tC²/ha², the empirical bias was insignificant for all sample sizes except for $n = 5$ (for SRE-corr). For a SOC stock variance of 2,100 tC²/ha², the empirical bias was insignificant for all sample sizes except for $n = 80$ (for all estimators), and additionally for $n = 10$ and $n = 20$ (for SRE-corr). Figure 1 also shows that, as the sample size increases, the empirical biases for SRE and HTE converge to zero. There is no visually noticeable difference in the convergence behaviours of these estimators.

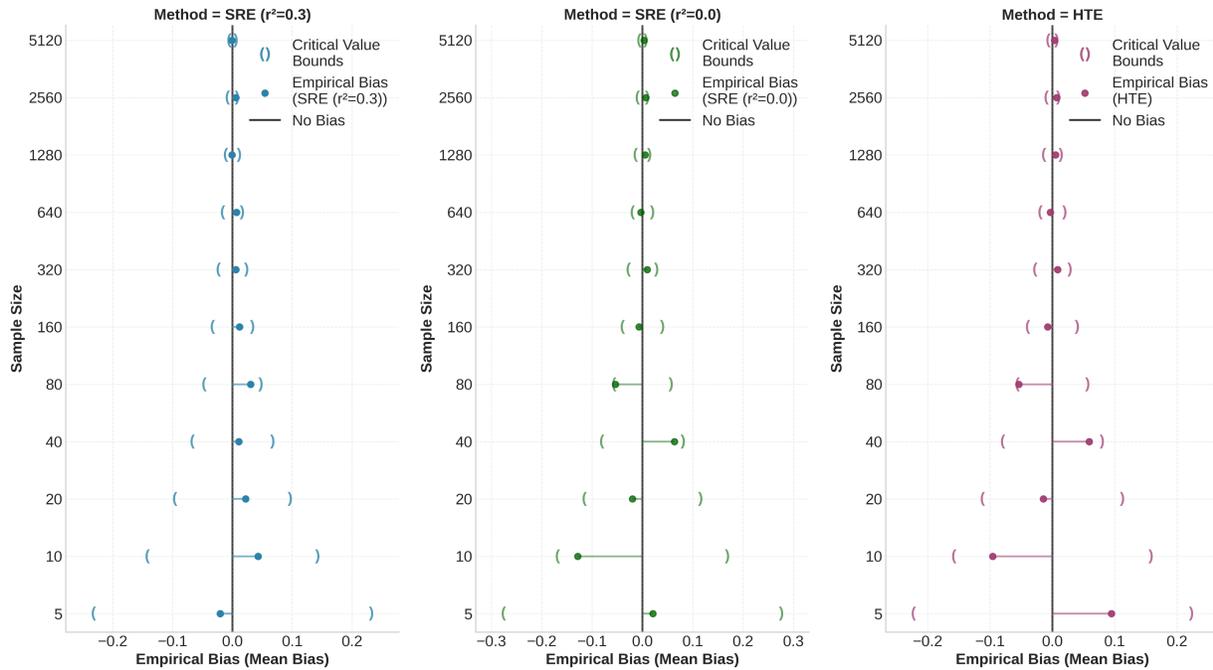

a) $Var(Z)$ = 500 tC²/ha²

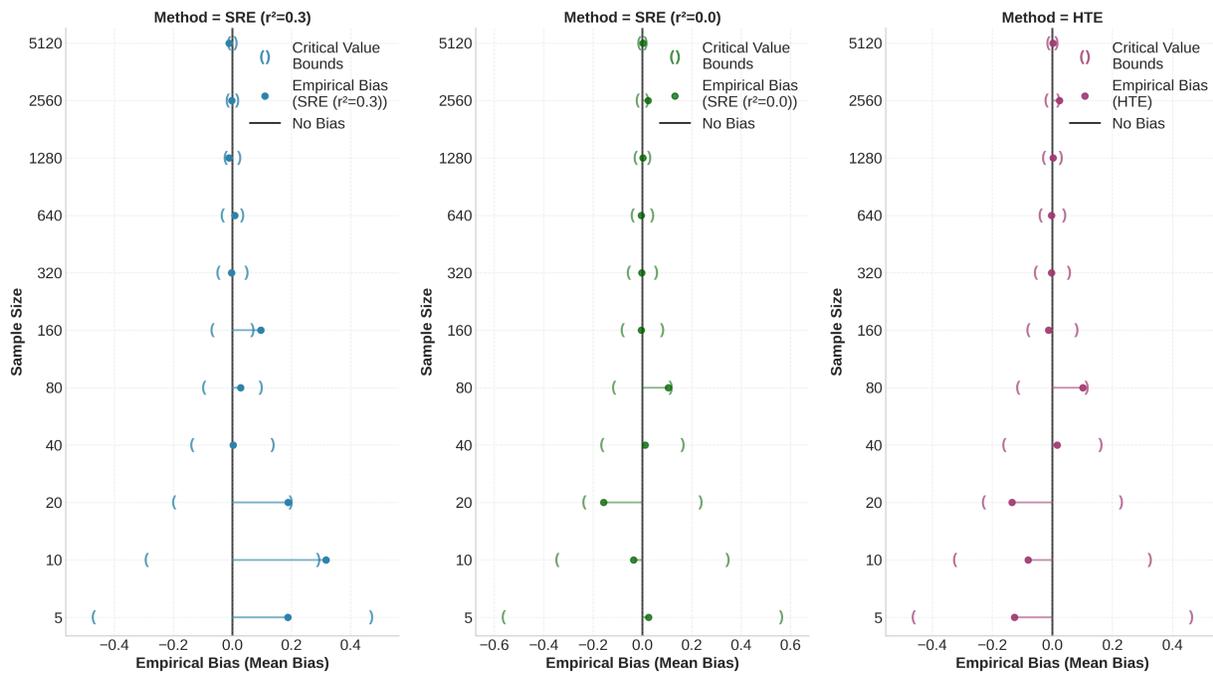

b) $Var(Z)$ = 2,100 tC²/ha²

**Figure 1.** Empirical biases along with critical values for the t-test ($\alpha$ = 5%) for SRE and HTE under various simulated sample sizes, for SOC stock variances of (a) 500 tC²/ha² and (b) 2,100 tC²/ha². Refer to Figures 3a-f in Appendix II for the full set of simulated population variances.

## 4.2. Empirical coverage probability

The simulation results depicted in Figure 2 reveal that the estimated confidence intervals for all estimators achieve the nominal coverage probability of 95% starting from a sample size of $n \geq 40$, regardless of SOC stock variance. For smaller sample sizes, the empirical coverage probabilities were substantially lower than the nominal value. This indicates that the approximative sampling variance estimator for SRE underestimates uncertainties at small sample sizes. Interestingly, the confidence intervals estimated for HTE were also affected by undercoverage, although to a lower degree. This observation suggests that at least part of the undercoverage of SRE could be explained by an unmet normality assumption (at small sample sizes) when constructing the confidence intervals via t-values.

Quantitatively, empirical coverage probabilities range from 73% with $n = 5$ to about 94% when $n = 40$, and then stabilize at 95% (±1%) for $n > 40$. The convergence to the nominal coverage probability confirms the asymptotic validity of the approximate sampling variance estimator and the confidence interval construction approach. The key observation for practitioners is that the empirical coverage probability stabilizes consistently at $n = 40$ regardless of the SOC stock variance and regardless of whether a correlated or an uncorrelated ancillary variable is used for SRE.

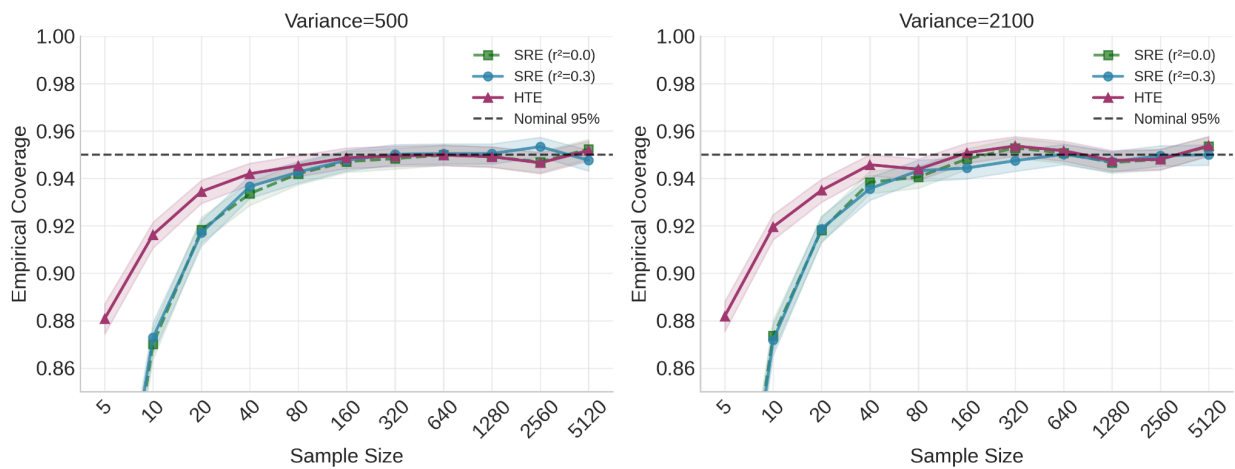

**Figure 2.** Empirical coverage probabilities for all estimators (SRE-uncorr, SRE-corr, and HTE) along with their MC uncertainties (y-axis) under various sample sizes (x-axis) and population variances of 500 tC²/ha² and 2,100 tC²/ha². The horizontal baseline indicates the nominal coverage probability of 95%. Refer to Figures 4 in Appendix II for the full set of simulated population variances.

## 4.3. Precision gain

Figure 3 depicts the precision gains achieved in our simulations when comparing SRE to HTE. If SRE is used with a correlated ancillary variable (SRE-corr with $r^2 \approx 0.3$), it consistently yields more precise estimates than HTE for sample sizes greater than 5. The plots clearly show that the precision gain converges to $(1 - r^2)$ around a sample size of 40, irrespective of the SOC stock variance. For instance, for a SOC stock variance of 2,100 tC²/ha² and a sample size of 1,280, SRE-corr has a sampling variance of 1.152, while HTE has a sampling variance of 1.671, yielding a precision gain of 0.7. This translates to SRE-corr being 30% more precise than HTE. This empirical observation confirms the theoretical relationship between the precision gain and the correlation of the ancillary variable and SOC stock, as formalized in Equation 9.

When using an uncorrelated ancillary variable (SRE-uncorr), no precision gains over HTE can be expected. In fact, the simulation results show that for small sample sizes < 20, SRE-uncorr may lead to less precise estimates than HTE. This observation is consistent across all SOC stock variances. These results imply that with a poor working model and limited sample data, SRE can be less efficient than HTE. However, as the sample size increases, the precision gain for SRE-uncorr rapidly approaches 1. Starting from a sample size of 40, there is no relevant difference in precision between SRE-uncorr and HTE. Therefore, the risk of obtaining an estimate with a lower precision than HTE can easily be controlled by using SRE only for sample sizes larger than 40.

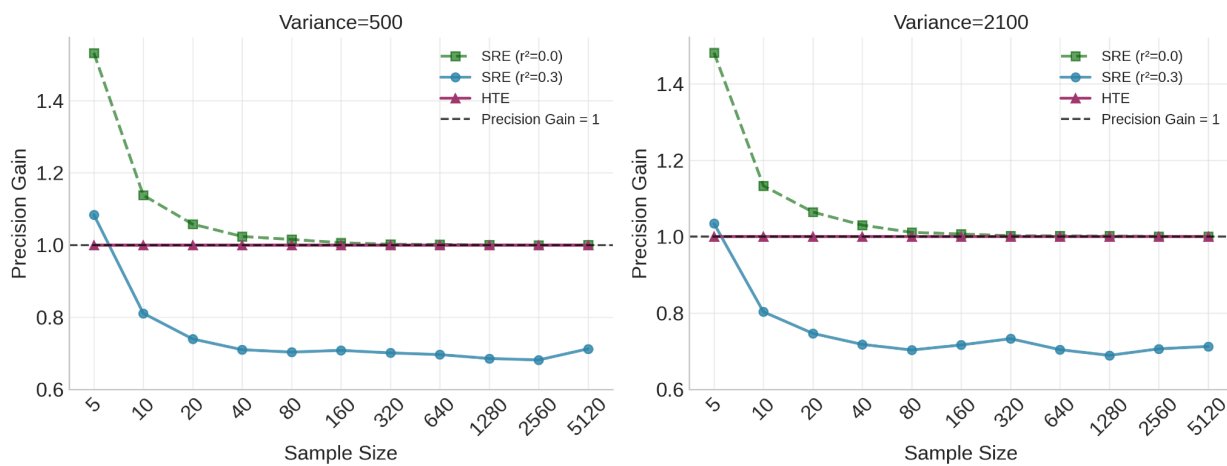

**Figure 3.** Precision gain (y-axis) under various sample sizes (x-axis) and population variances of 500 tC²/ha² and 2,100 tC²/ha². Comparison is made across all simulated estimators, i.e., SRE-uncorr, SRE-corr, and the HT estimator. Refer to Figures 5 in Appendix II for the full set of simulated population variances.

# 5. Synthesis

The goal of the above simulation study was to assess whether model-assisted estimators can provide higher precision SOC stock (change) estimates without compromising on scientific integrity. The findings provide evidence that the relationship holds true, provided that the minimum sample size criterion is satisfied and a correlated ancillary variable is incorporated.

## 5.1. Minimum required sample size for scientific integrity

The simulations show that SOC stock mean estimates obtained from SRE are scientifically sound if a minimum sample size of $n \geq 40$ is used. Starting from this point, the bias of the estimates is negligibly small, and their confidence intervals have the correct coverage probabilities, regardless of the SOC stock variance in the project area. Thus, applying SRE with $n \geq 40$ samples does not jeopardize the scientific integrity of a SOC sequestration project. Below a sample size of 40, estimates obtained from SRE are not reliable, as they may suffer from poorly calibrated confidence intervals.

We stress that this is a minimum requirement for **scientific integrity**, not for **practical utility** or **economic efficiency** of a SOC sequestration project: While sample sizes of 40 consistently yield unbiased estimates and correctly calibrated confidence intervals in our simulations, the associated uncertainties can be too large to detect or measure SOC stock changes over short time periods. In practice, larger sample sizes than 40 are typically required to reduce estimation uncertainties down to a level that allows measuring SOC sequestration due to agricultural land management changes, in particular, over relatively short time periods (e.g., 5 years). Ultimately, sample sizes for SOC sequestration projects should be determined by a power analysis based on the anticipated SOC sequestration rates, or by trading off sampling costs and uncertainty deductions to find an economically optimum number of samples (EONS) for a given SOC sequestration project.

Our findings are in line with existing literature which notes that "model-assisted estimators are asymptotically design-unbiased and design-consistent, irrespective of whether or not the working model is correctly specified" ([Dagdoug et al. 2021](#)). In their simulation study, the authors similarly confirmed that even more complex model-assisted estimators, e.g., based on random forest models, achieve near-zero bias and valid coverage probabilities for moderate sample sizes ([ibid.](#)). Experimental results like these and ours confirm the scientific integrity of model-assisted estimators.

## 5.2. Precision gains and project economics

Our simulations confirm that model-assisted estimators can reduce SOC stock estimation uncertainties, if the utilized ancillary variables have predictive power for SOC stocks. The empirical precision gains achieved by SRE are proportional to the squared correlation between the ancillary variable and SOC stock, which is in line with the theoretical results (c.f. Equation 9). The precision gains for SOC stock estimates directly propagate to precision gains for SOC stock change estimates and precision gains for effect size estimates (project change minus baseline change).[6] The latter is particularly important for SOC sequestration projects, as the precision of effect size estimates ultimately determines the uncertainty deductions to be applied before carbon credits are issued.

If the ancillary variables used for model-assisted estimation have no predictive power for SOC stocks, no precision gains can be expected. The simulations show that SRE may even yield estimates with a lower precision than a design-based estimation approach (HTE) if an uncorrelated ancillary variable ($r^2 = 0.0$) is used. However, inflated uncertainties only occurred in scenarios with very small sample sizes. For sample sizes $n \geq 40$ (as recommended for scientific integrity), the model-assisted estimator yields estimates with exactly the same level of uncertainty as the design-based estimator. The risk of obtaining estimates with inflated uncertainties can thus be managed easily by making sure the sample size is not too small. In any case, we recommend that project proponents employing a model-assisted estimator always additionally compute a design-based estimate from the sample data to be used for comparison and as a fallback option.

Overall, the application of SRE or other model-assisted estimators in a SOC sequestration project has the potential to **reduce sampling costs** (by maintaining the same level of precision with fewer samples) or **reduce uncertainty deductions**, or combinations of both. Sampling costs and uncertainty deductions can be pivotal in determining the difference between economically viable and unviable projects. Due to the maintained scientific integrity of SRE (for $n \geq 40$), model-assisted estimation is a straightforward way to improve project economics without compromising the credibility of carbon credits.

---

[6] For independent sampling designs, the SOC stock change estimation uncertainty (sampling variance) is simply the sum of the individual SOC stock estimation uncertainties. If the latter are reduced by a certain factor, the former is reduced by the same factor. For paired sampling designs, the covariances between the SOC stock estimates need to be taken into account. Analogously, effect size estimation uncertainties are sums of the SOC stock change estimation uncertainties.

# 6. Conclusion

This study demonstrates that model-assisted estimation offers a compelling value proposition as a tool for SOC MRV and the broader carbon markets. Methodological frameworks that do not explicitly permit model-assisted estimation, such as [VCS Methodology VM0042 v2.1 (2024)](#) and the EU CRCF draft, create economic disadvantages for project proponents due to avoidable uncertainty deductions and/or unnecessary sampling costs. Based on our findings, we offer the following recommendations:

1. **For regulatory bodies:** Methodological frameworks should be updated to explicitly permit model-assisted estimators and provide clear guidance for their correct implementation to ensure the highest levels of scientific integrity. In particular, a minimum sample size must be enforced so that the asymptotic properties of model-assisted estimators are practically realized. For SRE, our findings suggest a minimum sample size of $n \geq 40$ as a lower bound for scientific integrity.
2. **For project proponents:** Model-assisted estimators (e.g., SRE) should be the preferred option over design-based estimators whenever a correlated ancillary variable/models are available for the project area. Model-assisted estimators leverage the correlation between the ancillary variable and SOC stocks and can substantially improve project economics.
3. **For the SOC MRV industry:** Investment should be directed toward identifying and developing high-quality ancillary variables for SOC stock estimation. For example, there is a pressing need for Digital Soil Maps (DSMs) that are strongly correlated with SOC stock at the scale of a typical SOC sequestration project (i.e., regional clusters of farms), as globally calibrated DSMs often exhibit low correlations at the field- or farm-level. The highest precision gains, and thus the best economic outcomes, will be realized from bespoke DSMs that are fine-tuned to the specific project area.

# Appendix I

Table 1. Empirical bias one-sample two-tailed t-test across all simulated realities and estimators, SRE ($r^2 = 0$), SRE ($r^2 = 0.3$), and HTE.

| population variance | r² score | sample size | estimator | empirical bias | t statistic | p value | statistically significant |
|---|---|---|---|---|---|---|---|
| 100 | 0 | 5 | HTE | -0.025 | -0.550 | 0.5824 | FALSE |
| 100 | 0 | 5 | SRE | 0.024 | 0.446 | 0.6554 | FALSE |
| 100 | 0.3 | 5 | SRE | 0.037 | 0.806 | 0.4200 | FALSE |
| 100 | 0 | 10 | HTE | -0.020 | -0.630 | 0.5287 | FALSE |
| 100 | 0 | 10 | SRE | -0.033 | -0.980 | 0.3271 | FALSE |
| 100 | 0.3 | 10 | SRE | -0.012 | -0.416 | 0.6773 | FALSE |
| 100 | 0 | 20 | HTE | -0.003 | -0.144 | 0.8858 | FALSE |
| 100 | 0 | 20 | SRE | 0.000 | -0.009 | 0.9929 | FALSE |
| 100 | 0.3 | 20 | SRE | -0.007 | -0.355 | 0.7229 | FALSE |
| 100 | 0 | 40 | HTE | -0.011 | -0.688 | 0.4912 | FALSE |
| 100 | 0 | 40 | SRE | -0.010 | -0.638 | 0.5237 | FALSE |
| 100 | 0.3 | 40 | SRE | -0.020 | -1.478 | 0.1396 | FALSE |
| 100 | 0 | 80 | HTE | -0.013 | -1.202 | 0.2294 | FALSE |
| 100 | 0 | 80 | SRE | -0.014 | -1.261 | 0.2072 | FALSE |
| 100 | 0.3 | 80 | SRE | 0.009 | 0.990 | 0.3224 | FALSE |
| 100 | 0 | 160 | HTE | 0.003 | 0.364 | 0.7160 | FALSE |
| 100 | 0 | 160 | SRE | 0.003 | 0.390 | 0.6963 | FALSE |
| 100 | 0.3 | 160 | SRE | 0.006 | 0.859 | 0.3906 | FALSE |
| 100 | 0 | 320 | HTE | 0.003 | 0.550 | 0.5826 | FALSE |

| population variance | r² score | sample size | estimator | empirical bias | t statistic | p value | statistically significant |
|---|---|---|---|---|---|---|---|
| 100 | 0 | 320 | SRE | 0.003 | 0.503 | 0.6152 | FALSE |
| 100 | 0.3 | 320 | SRE | -0.001 | -0.252 | 0.8009 | FALSE |
| 100 | 0 | 640 | HTE | 0.005 | 1.310 | 0.1903 | FALSE |
| 100 | 0 | 640 | SRE | 0.005 | 1.319 | 0.1872 | FALSE |
| 100 | 0.3 | 640 | SRE | 0.001 | 0.208 | 0.8353 | FALSE |
| 100 | 0 | 1280 | HTE | -0.003 | -1.115 | 0.2648 | FALSE |
| 100 | 0 | 1280 | SRE | -0.003 | -1.103 | 0.2700 | FALSE |
| 100 | 0.3 | 1280 | SRE | 0.002 | 0.689 | 0.4906 | FALSE |
| 100 | 0 | 2560 | HTE | -0.001 | -0.655 | 0.5123 | FALSE |
| 100 | 0 | 2560 | SRE | -0.001 | -0.621 | 0.5344 | FALSE |
| 100 | 0.3 | 2560 | SRE | 0.000 | -0.210 | 0.8340 | FALSE |
| 100 | 0 | 5120 | HTE | -0.001 | -0.437 | 0.6621 | FALSE |
| 100 | 0 | 5120 | SRE | -0.001 | -0.448 | 0.6543 | FALSE |
| 100 | 0.3 | 5120 | SRE | -0.002 | -1.644 | 0.1003 | FALSE |
| 500 | 0 | 5 | HTE | 0.095 | 0.954 | 0.3401 | FALSE |
| 500 | 0 | 5 | SRE | 0.021 | 0.170 | 0.8650 | FALSE |
| 500 | 0.3 | 5 | SRE | -0.020 | -0.196 | 0.8446 | FALSE |
| 500 | 0 | 10 | HTE | -0.096 | -1.357 | 0.1750 | FALSE |
| 500 | 0 | 10 | SRE | -0.128 | -1.707 | 0.0878 | FALSE |
| 500 | 0.3 | 10 | SRE | 0.043 | 0.676 | 0.4993 | FALSE |
| 500 | 0 | 20 | HTE | -0.015 | -0.291 | 0.7711 | FALSE |
| 500 | 0 | 20 | SRE | -0.020 | -0.387 | 0.6991 | FALSE |
| 500 | 0.3 | 20 | SRE | 0.022 | 0.509 | 0.6111 | FALSE |

| population variance | r² score | sample size | estimator | empirical bias | t statistic | p value | statistically significant |
|---|---|---|---|---|---|---|---|
| 500 | 0 | 40 | HTE | 0.059 | 1.662 | 0.0965 | FALSE |
| 500 | 0 | 40 | SRE | 0.064 | 1.769 | 0.0770 | FALSE |
| 500 | 0.3 | 40 | SRE | 0.011 | 0.356 | 0.7216 | FALSE |
| 500 | 0.3 | 80 | SRE | 0.030 | 1.432 | 0.1522 | FALSE |
| 500 | 0 | 80 | HTE | -0.054 | -2.130 | 0.033 | TRUE |
| 500 | 0 | 80 | SRE | -0.054 | -2.114 | 0.035 | TRUE |
| 500 | 0 | 160 | HTE | -0.008 | -0.437 | 0.6618 | FALSE |
| 500 | 0 | 160 | SRE | -0.007 | -0.379 | 0.7049 | FALSE |
| 500 | 0.3 | 160 | SRE | 0.012 | 0.799 | 0.4242 | FALSE |
| 500 | 0 | 320 | HTE | 0.008 | 0.671 | 0.5022 | FALSE |
| 500 | 0 | 320 | SRE | 0.010 | 0.773 | 0.4396 | FALSE |
| 500 | 0.3 | 320 | SRE | 0.006 | 0.566 | 0.5712 | FALSE |
| 500 | 0 | 640 | HTE | -0.003 | -0.365 | 0.7149 | FALSE |
| 500 | 0 | 640 | SRE | -0.003 | -0.368 | 0.7131 | FALSE |
| 500 | 0.3 | 640 | SRE | 0.007 | 0.936 | 0.3492 | FALSE |
| 500 | 0 | 1280 | HTE | 0.005 | 0.841 | 0.4005 | FALSE |
| 500 | 0 | 1280 | SRE | 0.005 | 0.847 | 0.3970 | FALSE |
| 500 | 0.3 | 1280 | SRE | -0.001 | -0.113 | 0.9102 | FALSE |
| 500 | 0 | 2560 | HTE | 0.007 | 1.541 | 0.1234 | FALSE |
| 500 | 0 | 2560 | SRE | 0.007 | 1.577 | 0.1149 | FALSE |
| 500 | 0.3 | 2560 | SRE | 0.006 | 1.614 | 0.1066 | FALSE |
| 500 | 0 | 5120 | HTE | 0.003 | 1.057 | 0.2905 | FALSE |
| 500 | 0 | 5120 | SRE | 0.003 | 1.086 | 0.2776 | FALSE |

| population variance | r² score | sample size | estimator | empirical bias | t statistic | p value | statistically significant |
|---|---|---|---|---|---|---|---|
| 500 | 0.3 | 5120 | SRE | -0.001 | -0.306 | 0.7600 | FALSE |
| 900 | 0 | 5 | HTE | 0.178 | 1.322 | 0.1862 | FALSE |
| 900 | 0 | 5 | SRE | 0.216 | 1.311 | 0.1898 | FALSE |
| 900 | 0.3 | 5 | SRE | -0.307 | -2.118 | 0.034 | TRUE |
| 900 | 0 | 10 | HTE | 0.069 | 0.731 | 0.4646 | FALSE |
| 900 | 0 | 10 | SRE | 0.046 | 0.456 | 0.6487 | FALSE |
| 900 | 0.3 | 10 | SRE | -0.121 | -1.432 | 0.1522 | FALSE |
| 900 | 0 | 20 | HTE | 0.077 | 1.167 | 0.2434 | FALSE |
| 900 | 0 | 20 | SRE | 0.078 | 1.140 | 0.2545 | FALSE |
| 900 | 0.3 | 20 | SRE | -0.039 | -0.671 | 0.5024 | FALSE |
| 900 | 0 | 40 | HTE | 0.045 | 0.951 | 0.3416 | FALSE |
| 900 | 0 | 40 | SRE | 0.043 | 0.904 | 0.3661 | FALSE |
| 900 | 0.3 | 40 | SRE | -0.022 | -0.553 | 0.5803 | FALSE |
| 900 | 0 | 80 | HTE | 0.009 | 0.262 | 0.7934 | FALSE |
| 900 | 0 | 80 | SRE | 0.011 | 0.311 | 0.7556 | FALSE |
| 900 | 0.3 | 80 | SRE | -0.043 | -1.513 | 0.1302 | FALSE |
| 900 | 0 | 160 | HTE | -0.001 | -0.024 | 0.9809 | FALSE |
| 900 | 0 | 160 | SRE | -0.001 | -0.022 | 0.9825 | FALSE |
| 900 | 0.3 | 160 | SRE | 0.014 | 0.706 | 0.4805 | FALSE |
| 900 | 0 | 320 | HTE | 0.008 | 0.488 | 0.6256 | FALSE |
| 900 | 0 | 320 | SRE | 0.007 | 0.431 | 0.6667 | FALSE |
| 900 | 0.3 | 320 | SRE | -0.014 | -1.014 | 0.3106 | FALSE |
| 900 | 0 | 640 | HTE | -0.002 | -0.162 | 0.8713 | FALSE |

| population variance | r² score | sample size | estimator | empirical bias | t statistic | p value | statistically significant |
|---|---|---|---|---|---|---|---|
| 900 | 0 | 640 | SRE | -0.002 | -0.129 | 0.8973 | FALSE |
| 900 | 0.3 | 640 | SRE | 0.010 | 1.054 | 0.2920 | FALSE |
| 900 | 0 | 1280 | HTE | -0.014 | -1.709 | 0.0874 | FALSE |
| 900 | 0 | 1280 | SRE | -0.014 | -1.722 | 0.0851 | FALSE |
| 900 | 0.3 | 1280 | SRE | 0.002 | 0.226 | 0.8209 | FALSE |
| 900 | 0 | 2560 | HTE | 0.000 | -0.084 | 0.9331 | FALSE |
| 900 | 0 | 2560 | SRE | 0.000 | -0.061 | 0.9516 | FALSE |
| 900 | 0.3 | 2560 | SRE | -0.006 | -1.268 | 0.2049 | FALSE |
| 900 | 0 | 5120 | HTE | -0.005 | -1.100 | 0.2713 | FALSE |
| 900 | 0 | 5120 | SRE | -0.005 | -1.086 | 0.2774 | FALSE |
| 900 | 0.3 | 5120 | SRE | 0.000 | -0.084 | 0.9327 | FALSE |
| 1300 | 0 | 5 | HTE | 0.093 | 0.575 | 0.5656 | FALSE |
| 1300 | 0 | 5 | SRE | 0.144 | 0.729 | 0.4662 | FALSE |
| 1300 | 0.3 | 5 | SRE | -0.120 | -0.735 | 0.4624 | FALSE |
| 1300 | 0 | 10 | HTE | 0.021 | 0.187 | 0.8516 | FALSE |
| 1300 | 0 | 10 | SRE | 0.051 | 0.420 | 0.6742 | FALSE |
| 1300 | 0.3 | 10 | SRE | 0.001 | 0.010 | 0.9923 | FALSE |
| 1300 | 0 | 20 | HTE | 0.025 | 0.311 | 0.7561 | FALSE |
| 1300 | 0 | 20 | SRE | 0.036 | 0.434 | 0.6644 | FALSE |
| 1300 | 0.3 | 20 | SRE | -0.018 | -0.263 | 0.7927 | FALSE |
| 1300 | 0 | 40 | HTE | 0.025 | 0.435 | 0.6634 | FALSE |
| 1300 | 0 | 40 | SRE | 0.030 | 0.529 | 0.5968 | FALSE |
| 1300 | 0.3 | 40 | SRE | 0.033 | 0.674 | 0.5002 | FALSE |

| population variance | r² score | sample size | estimator | empirical bias | t statistic | p value | statistically significant |
|---|---|---|---|---|---|---|---|
| 1300 | 0 | 80 | HTE | -0.031 | -0.772 | 0.4399 | FALSE |
| 1300 | 0 | 80 | SRE | -0.034 | -0.823 | 0.4105 | FALSE |
| 1300 | 0.3 | 80 | SRE | 0.031 | 0.923 | 0.3560 | FALSE |
| 1300 | 0 | 160 | HTE | -0.012 | -0.406 | 0.6846 | FALSE |
| 1300 | 0 | 160 | SRE | -0.012 | -0.435 | 0.6639 | FALSE |
| 1300 | 0.3 | 160 | SRE | 0.045 | 1.895 | 0.0581 | FALSE |
| 1300 | 0 | 320 | HTE | -0.013 | -0.656 | 0.5120 | FALSE |
| 1300 | 0 | 320 | SRE | -0.014 | -0.674 | 0.5006 | FALSE |
| 1300 | 0.3 | 320 | SRE | -0.026 | -1.561 | 0.1186 | FALSE |
| 1300 | 0 | 640 | HTE | -0.008 | -0.560 | 0.5757 | FALSE |
| 1300 | 0 | 640 | SRE | -0.008 | -0.550 | 0.5825 | FALSE |
| 1300 | 0.3 | 640 | SRE | -0.009 | -0.750 | 0.4532 | FALSE |
| 1300 | 0 | 1280 | HTE | -0.012 | -1.248 | 0.2121 | FALSE |
| 1300 | 0 | 1280 | SRE | -0.012 | -1.240 | 0.2150 | FALSE |
| 1300 | 0.3 | 1280 | SRE | -0.002 | -0.227 | 0.8204 | FALSE |
| 1300 | 0 | 2560 | HTE | 0.001 | 0.198 | 0.8429 | FALSE |
| 1300 | 0 | 2560 | SRE | 0.002 | 0.232 | 0.8163 | FALSE |
| 1300 | 0.3 | 2560 | SRE | 0.001 | 0.128 | 0.8985 | FALSE |
| 1300 | 0 | 5120 | HTE | 0.001 | 0.223 | 0.8235 | FALSE |
| 1300 | 0 | 5120 | SRE | 0.001 | 0.222 | 0.8243 | FALSE |
| 1300 | 0.3 | 5120 | SRE | -0.006 | -1.386 | 0.1658 | FALSE |
| 1700 | 0 | 5 | HTE | -0.104 | -0.569 | 0.5691 | FALSE |
| 1700 | 0 | 5 | SRE | -0.071 | -0.318 | 0.7506 | FALSE |

| population variance | r² score | sample size | estimator | empirical bias | t statistic | p value | statistically significant |
|---|---|---|---|---|---|---|---|
| 1700 | 0.3 | 5 | SRE | -0.046 | -0.251 | 0.8019 | FALSE |
| 1700 | 0 | 10 | HTE | 0.073 | 0.562 | 0.5744 | FALSE |
| 1700 | 0 | 10 | SRE | 0.173 | 1.252 | 0.2104 | FALSE |
| 1700 | 0.3 | 10 | SRE | -0.128 | -1.097 | 0.2727 | FALSE |
| 1700 | 0 | 20 | HTE | 0.112 | 1.223 | 0.2213 | FALSE |
| 1700 | 0 | 20 | SRE | 0.135 | 1.426 | 0.1539 | FALSE |
| 1700 | 0.3 | 20 | SRE | -0.093 | -1.172 | 0.2412 | FALSE |
| 1700 | 0 | 40 | HTE | -0.069 | -1.064 | 0.2876 | FALSE |
| 1700 | 0 | 40 | SRE | -0.060 | -0.907 | 0.3643 | FALSE |
| 1700 | 0.3 | 40 | SRE | -0.021 | -0.374 | 0.7087 | FALSE |
| 1700 | 0 | 80 | HTE | -0.045 | -0.988 | 0.3230 | FALSE |
| 1700 | 0 | 80 | SRE | -0.045 | -0.983 | 0.3257 | FALSE |
| 1700 | 0.3 | 80 | SRE | -0.017 | -0.426 | 0.6699 | FALSE |
| 1700 | 0 | 160 | HTE | 0.003 | 0.082 | 0.9350 | FALSE |
| 1700 | 0 | 160 | SRE | 0.004 | 0.128 | 0.8981 | FALSE |
| 1700 | 0.3 | 160 | SRE | -0.050 | -1.825 | 0.0680 | FALSE |
| 1700 | 0 | 320 | HTE | -0.028 | -1.193 | 0.2327 | FALSE |
| 1700 | 0 | 320 | SRE | -0.029 | -1.230 | 0.2189 | FALSE |
| 1700 | 0.3 | 320 | SRE | 0.015 | 0.753 | 0.4512 | FALSE |
| 1700 | 0 | 640 | HTE | 0.006 | 0.363 | 0.7163 | FALSE |
| 1700 | 0 | 640 | SRE | 0.005 | 0.306 | 0.7595 | FALSE |
| 1700 | 0.3 | 640 | SRE | -0.016 | -1.168 | 0.2429 | FALSE |
| 1700 | 0 | 1280 | HTE | 0.001 | 0.086 | 0.9318 | FALSE |

| population variance | r² score | sample size | estimator | empirical bias | t statistic | p value | statistically significant |
|---|---|---|---|---|---|---|---|
| 1700 | 0 | 1280 | SRE | 0.001 | 0.067 | 0.9465 | FALSE |
| 1700 | 0.3 | 1280 | SRE | -0.011 | -1.111 | 0.2665 | FALSE |
| 1700 | 0 | 2560 | HTE | 0.010 | 1.204 | 0.2287 | FALSE |
| 1700 | 0 | 2560 | SRE | 0.010 | 1.200 | 0.2300 | FALSE |
| 1700 | 0.3 | 2560 | SRE | -0.008 | -1.125 | 0.2606 | FALSE |
| 1700 | 0 | 5120 | HTE | 0.001 | 0.232 | 0.8168 | FALSE |
| 1700 | 0 | 5120 | SRE | 0.001 | 0.232 | 0.8167 | FALSE |
| 1700 | 0.3 | 5120 | SRE | 0.003 | 0.616 | 0.5378 | FALSE |
| 2100 | 0 | 5 | HTE | -0.126 | -0.612 | 0.5404 | FALSE |
| 2100 | 0 | 5 | SRE | 0.025 | 0.098 | 0.9218 | FALSE |
| 2100 | 0.3 | 5 | SRE | 0.188 | 0.894 | 0.3711 | FALSE |
| 2100 | 0 | 10 | HTE | -0.081 | -0.557 | 0.5773 | FALSE |
| 2100 | 0 | 10 | SRE | -0.036 | -0.232 | 0.8168 | FALSE |
| 2100 | 0.3 | 10 | SRE | 0.317 | 2.438 | 0.015 | TRUE |
| 2100 | 0 | 20 | HTE | -0.135 | -1.317 | 0.1879 | FALSE |
| 2100 | 0 | 20 | SRE | -0.157 | -1.489 | 0.1365 | FALSE |
| 2100 | 0.3 | 20 | SRE | 0.189 | 2.135 | 0.033 | TRUE |
| 2100 | 0 | 40 | HTE | 0.016 | 0.226 | 0.8210 | FALSE |
| 2100 | 0 | 40 | SRE | 0.011 | 0.150 | 0.8807 | FALSE |
| 2100 | 0.3 | 40 | SRE | 0.002 | 0.038 | 0.9699 | FALSE |
| 2100 | 0.3 | 80 | SRE | 0.028 | 0.646 | 0.5183 | FALSE |
| 2100 | 0 | 80 | HTE | 0.101 | 1.972 | 0.049 | TRUE |
| 2100 | 0 | 80 | SRE | 0.105 | 2.022 | 0.043 | TRUE |

| population variance | r² score | sample size | estimator | empirical bias | t statistic | p value | statistically significant |
|---|---|---|---|---|---|---|---|
| 2100 | 0 | 160 | HTE | -0.013 | -0.348 | 0.7281 | FALSE |
| 2100 | 0 | 160 | SRE | -0.004 | -0.119 | 0.9055 | FALSE |
| 2100 | 0.3 | 160 | SRE | 0.096 | 3.156 | 0.002 | TRUE |
| 2100 | 0 | 320 | HTE | -0.003 | -0.104 | 0.9169 | FALSE |
| 2100 | 0 | 320 | SRE | -0.002 | -0.087 | 0.9303 | FALSE |
| 2100 | 0.3 | 320 | SRE | -0.003 | -0.129 | 0.8972 | FALSE |
| 2100 | 0 | 640 | HTE | -0.003 | -0.171 | 0.8644 | FALSE |
| 2100 | 0 | 640 | SRE | -0.005 | -0.266 | 0.7902 | FALSE |
| 2100 | 0.3 | 640 | SRE | 0.008 | 0.528 | 0.5977 | FALSE |
| 2100 | 0 | 1280 | HTE | 0.002 | 0.186 | 0.8525 | FALSE |
| 2100 | 0 | 1280 | SRE | 0.002 | 0.144 | 0.8856 | FALSE |
| 2100 | 0.3 | 1280 | SRE | -0.011 | -1.034 | 0.3010 | FALSE |
| 2100 | 0.3 | 2560 | SRE | -0.002 | -0.275 | 0.7833 | FALSE |
| 2100 | 0 | 2560 | HTE | 0.023 | 2.586 | 0.010 | TRUE |
| 2100 | 0 | 2560 | SRE | 0.023 | 2.577 | 0.010 | TRUE |
| 2100 | 0 | 5120 | HTE | 0.001 | 0.231 | 0.8173 | FALSE |
| 2100 | 0 | 5120 | SRE | 0.002 | 0.238 | 0.8118 | FALSE |
| 2100 | 0.3 | 5120 | SRE | -0.011 | -2.122 | 0.034 | TRUE |

# Appendix II

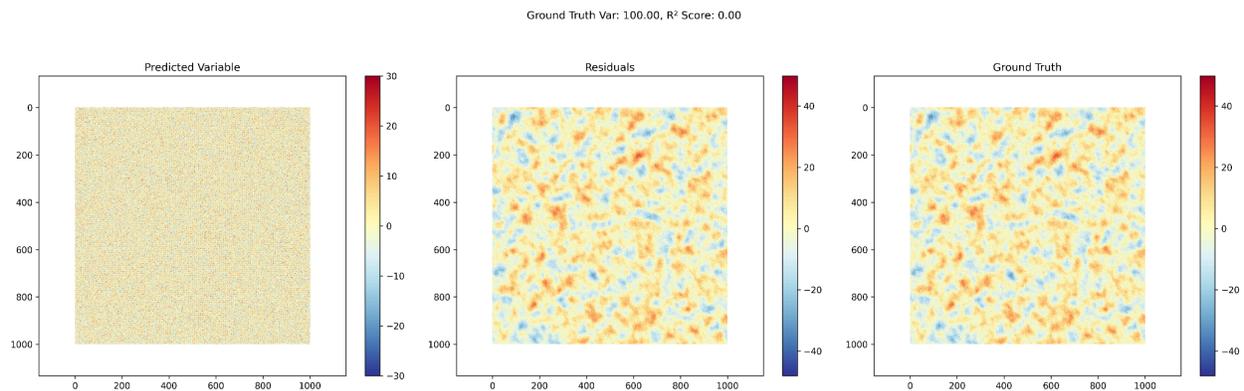

a)

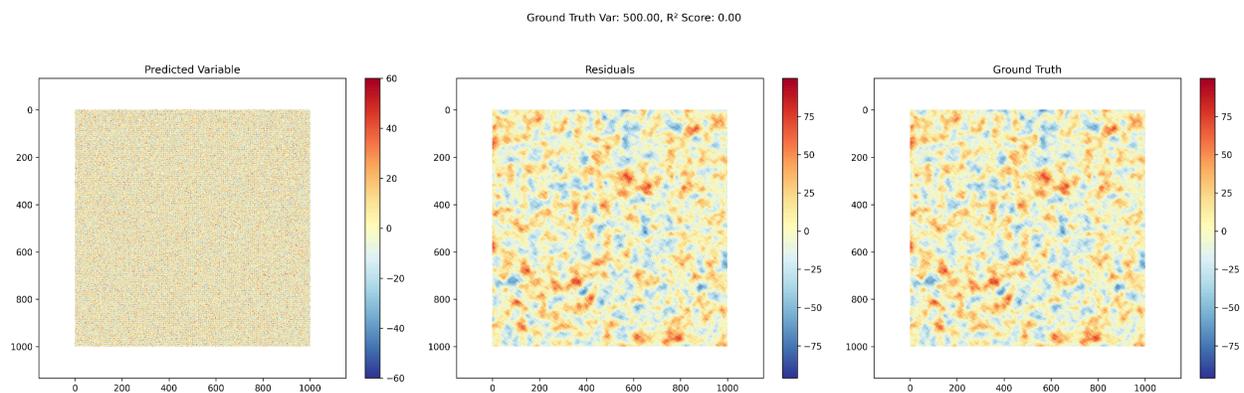

b)

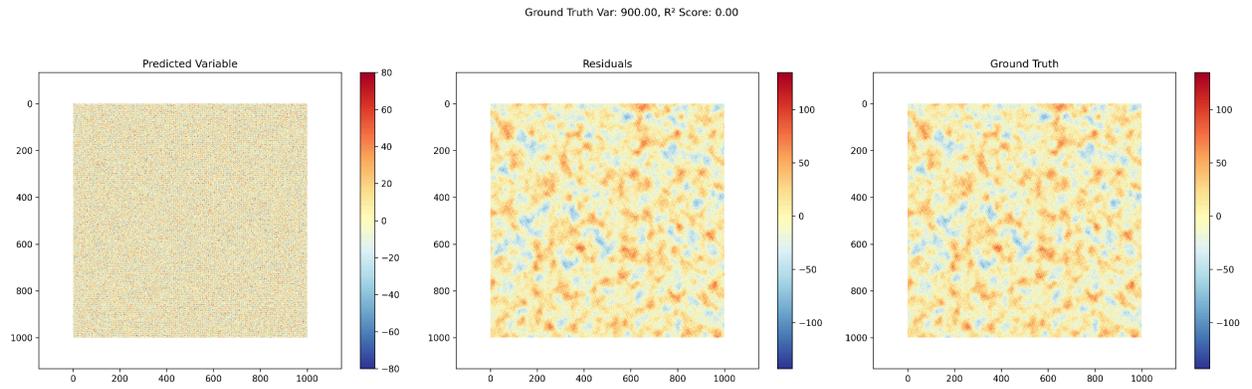

c)

**Figure 1.** Simulated predicted variable, residuals and ground truth population for a population variance ∈ {100, 500, 900, 1300, 1700, 2100}, true mean of 1 and predicted variable with R2 score of 0.

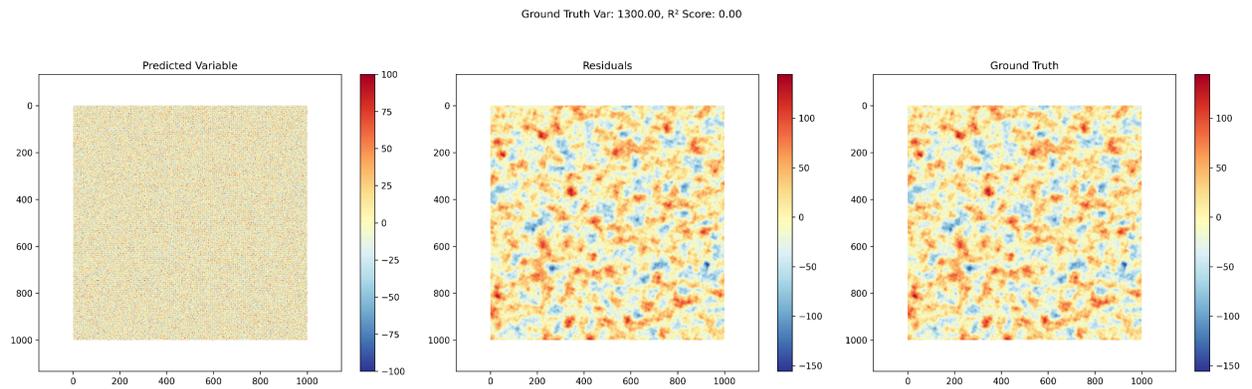

d)

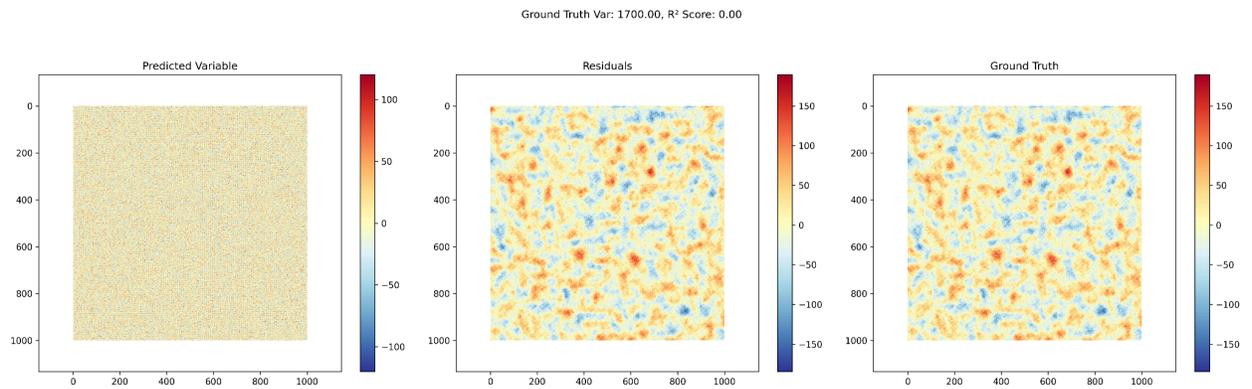

e)

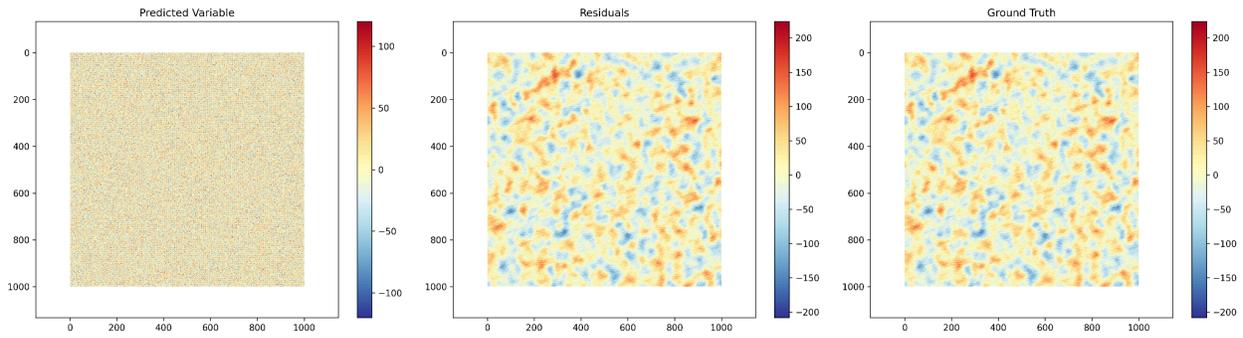

f)

**Figure 1.** Continued.

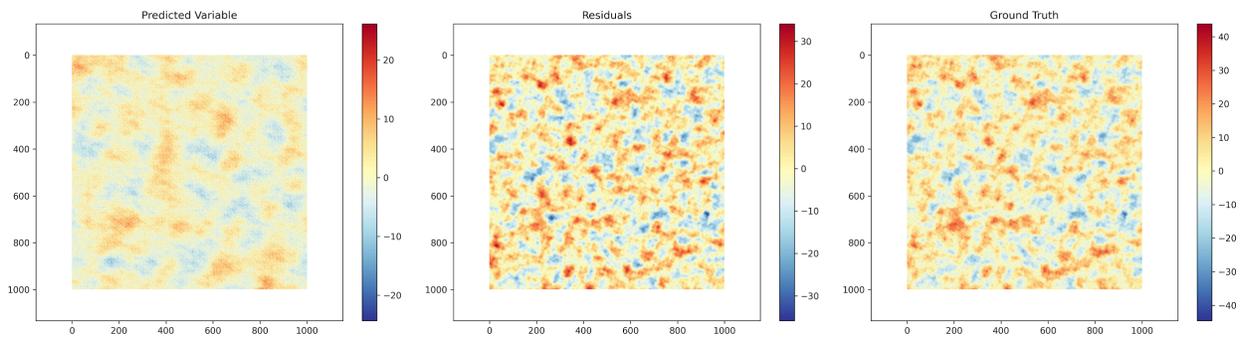

a)

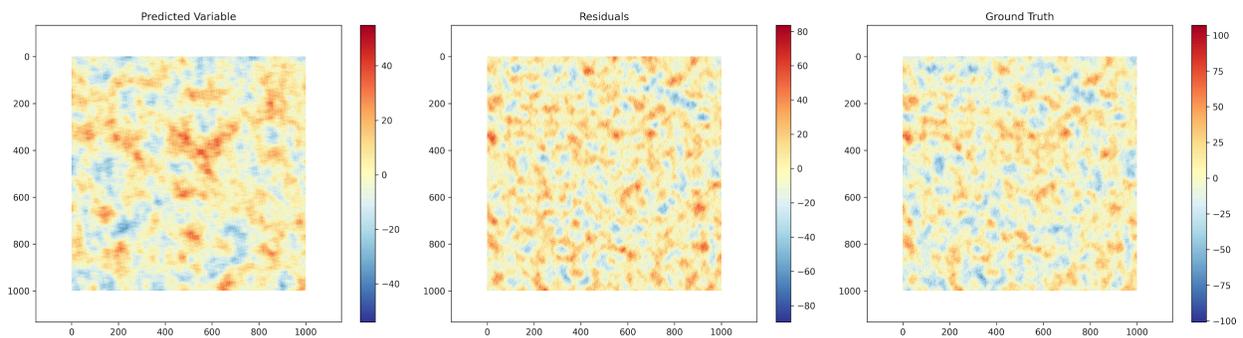

b)

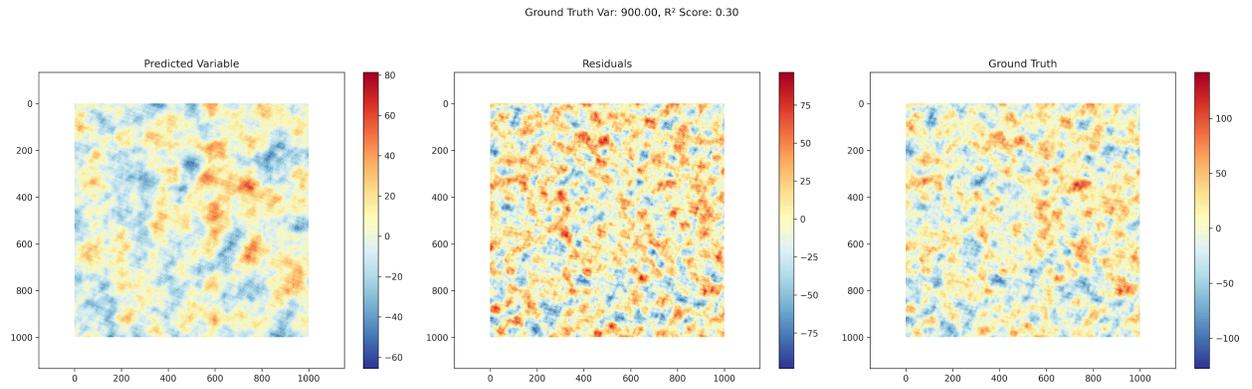

c)

**Figure 2.** Simulated predicted variable, residuals and ground truth population for a population variance ∈ {100, 500, 900, 1300, 1700, 2100}, true mean of 1 and predicted variable with R2 score of 0.3.

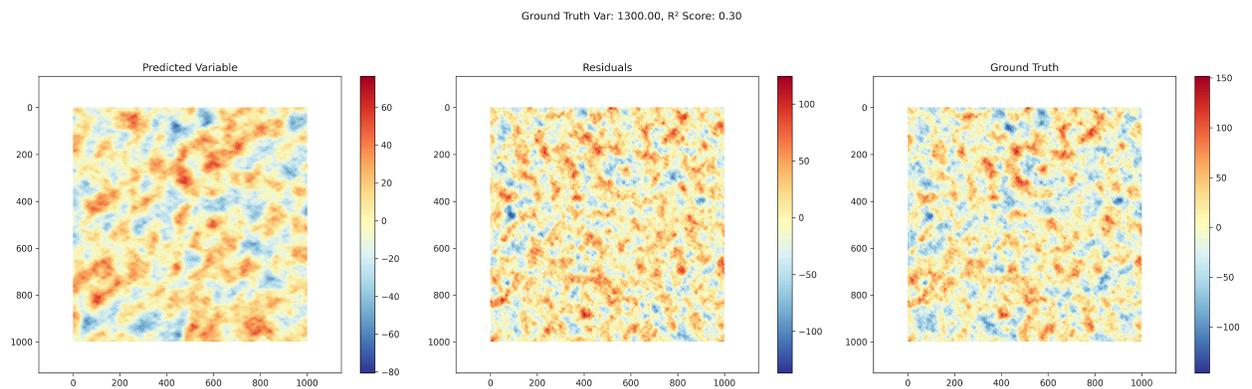

d)

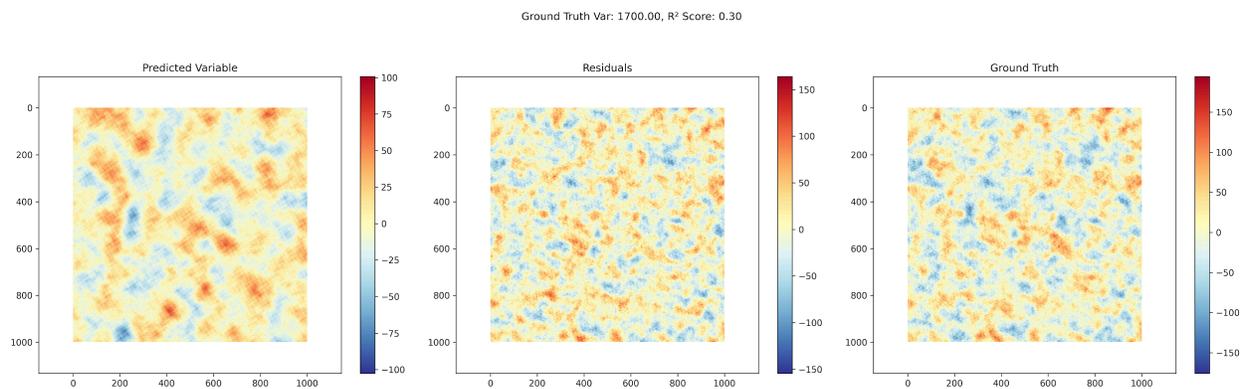

e)

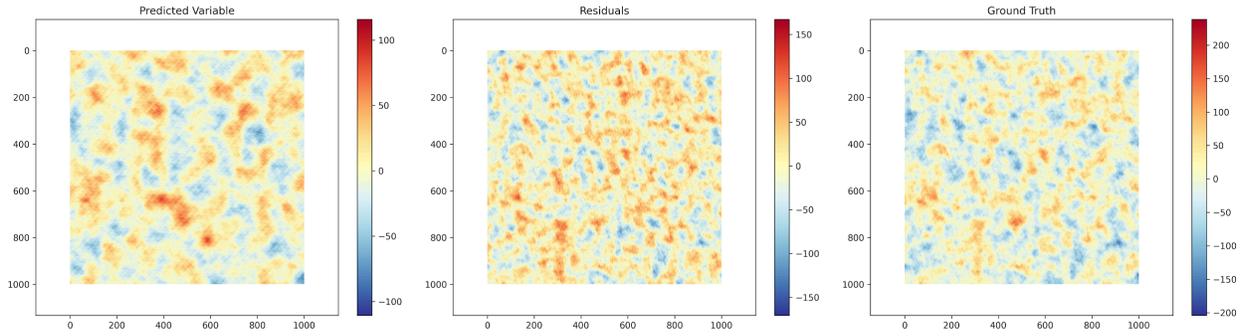

f)

**Figure 2.** Continued.

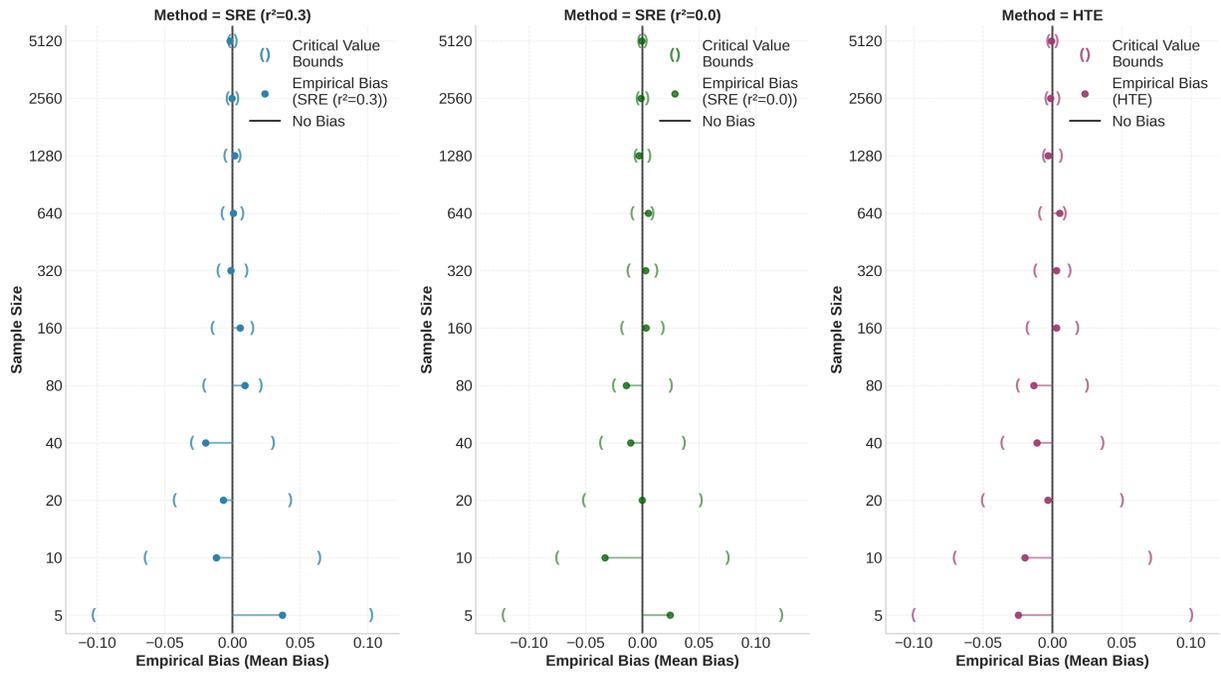

a)  $Var(Z)$ = 100 tC²/ha²

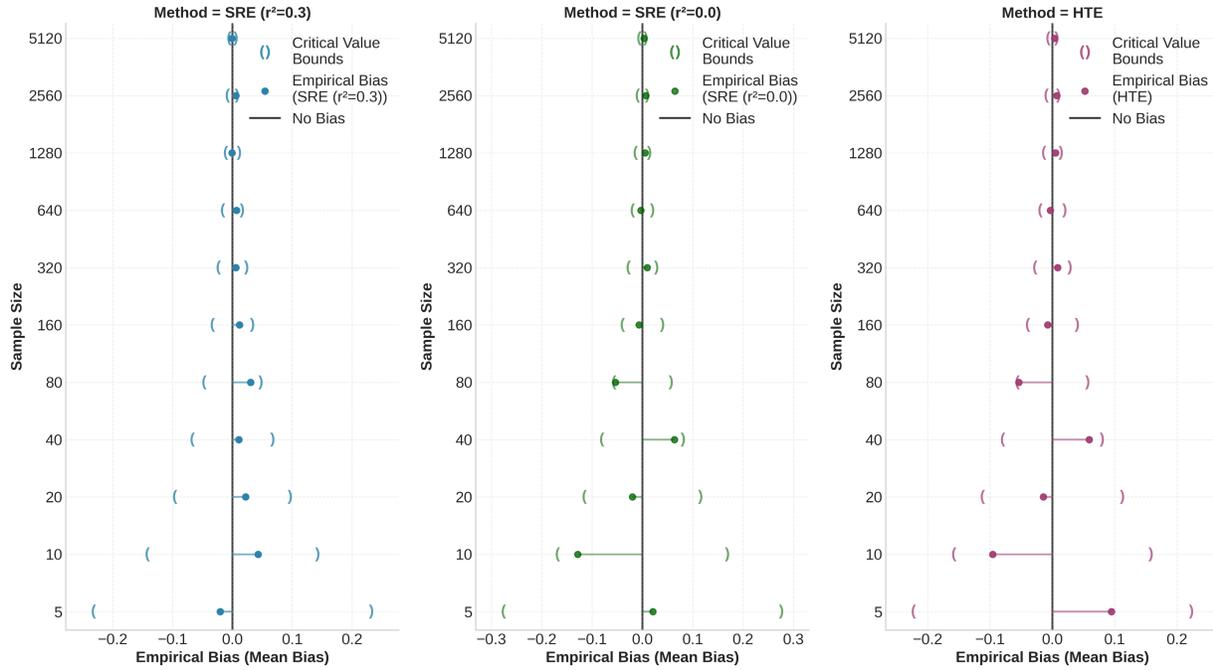

b) $Var(Z)$ = 500 tC²/ha²

**Figure 3.** Empirical biases along with critical values for the t-test (α = 5%) for SRE and HTE under various simulated sample sizes, for the 6 SOC stock simulated variances of (a) 100 tC²/ha², (b) 500 tC²/ha², (c) 900 tC²/ha², (d) 1300 tC²/ha², (e) 1700 tC²/ha², and (f) 2100 tC²/ha².

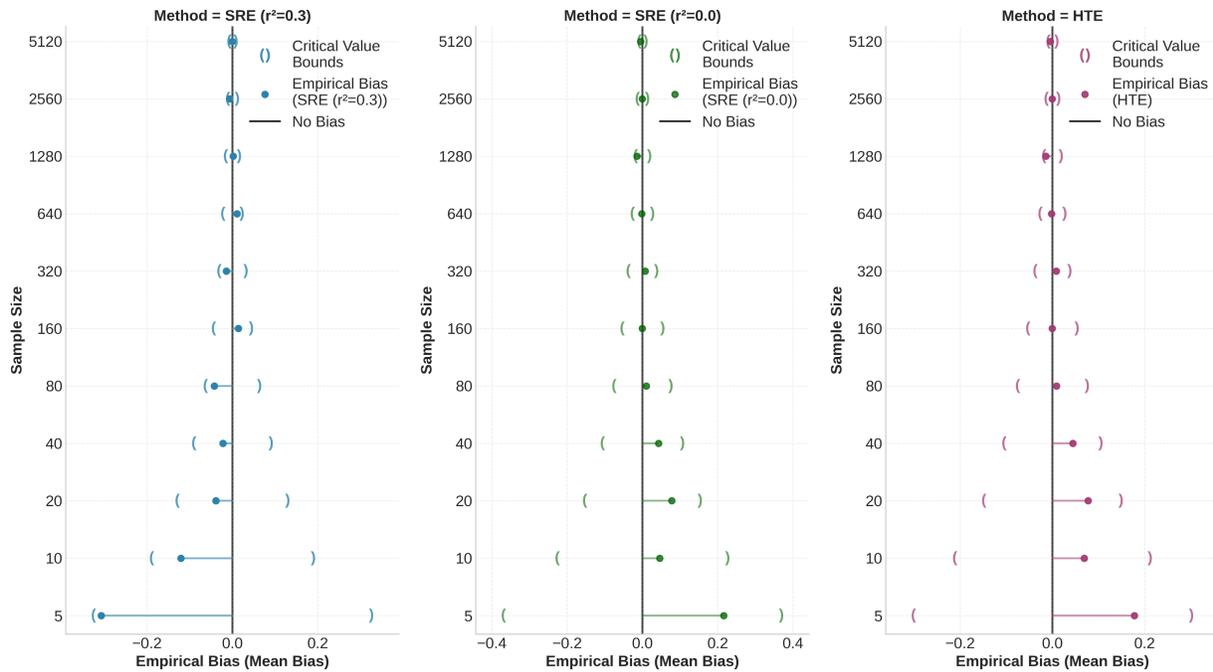

c) $Var(Z)$ = 900 tC²/ha²

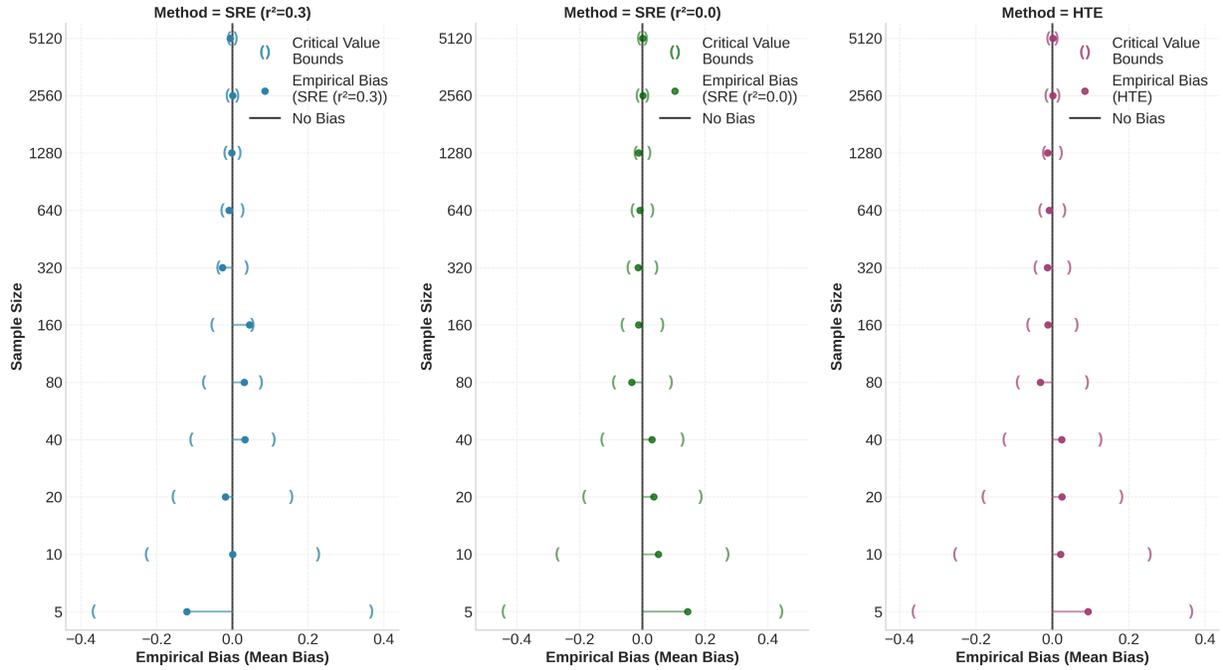

d) $Var(Z)$ = 1300 tC²/ha²

**Figure 3.** Continued.

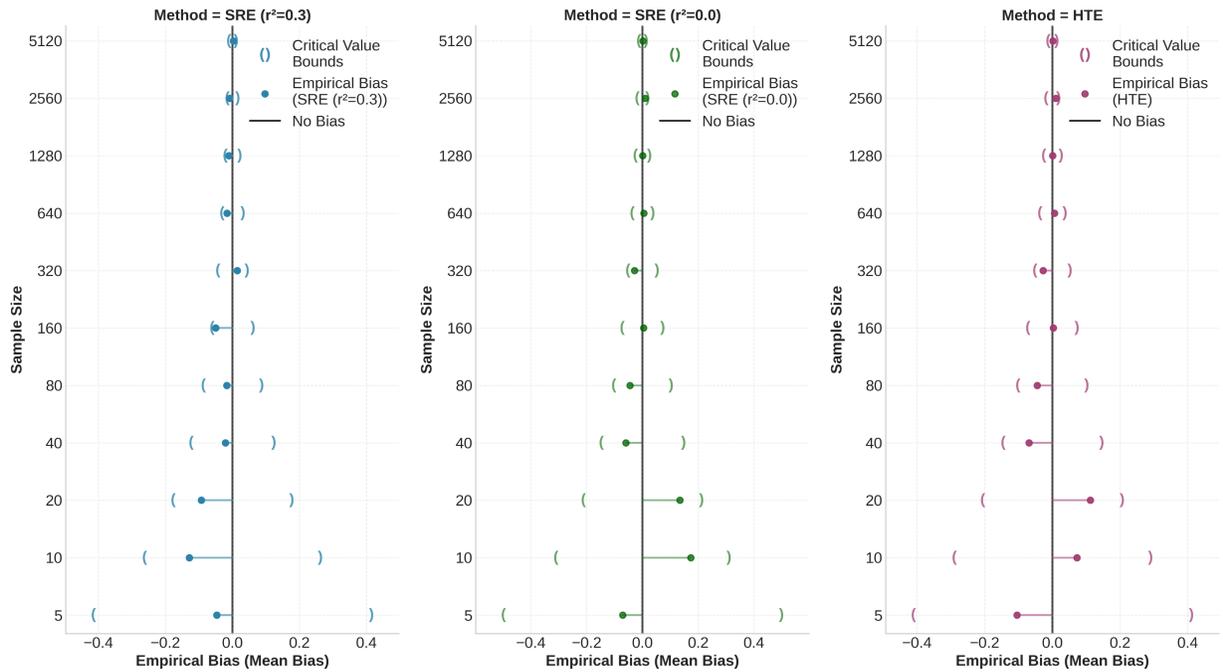

e) $Var(Z)$ = 1700 tC²/ha²

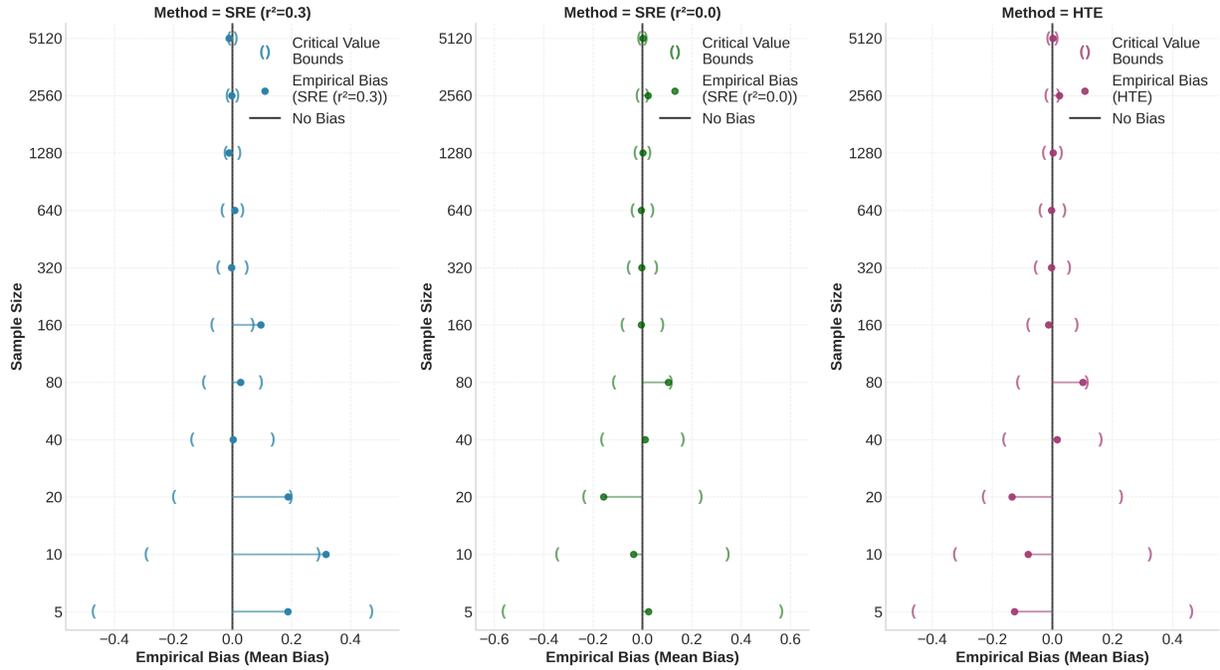

f) $Var(Z)$ = 2,100 tC²/ha²

**Figure 3.** Continued.

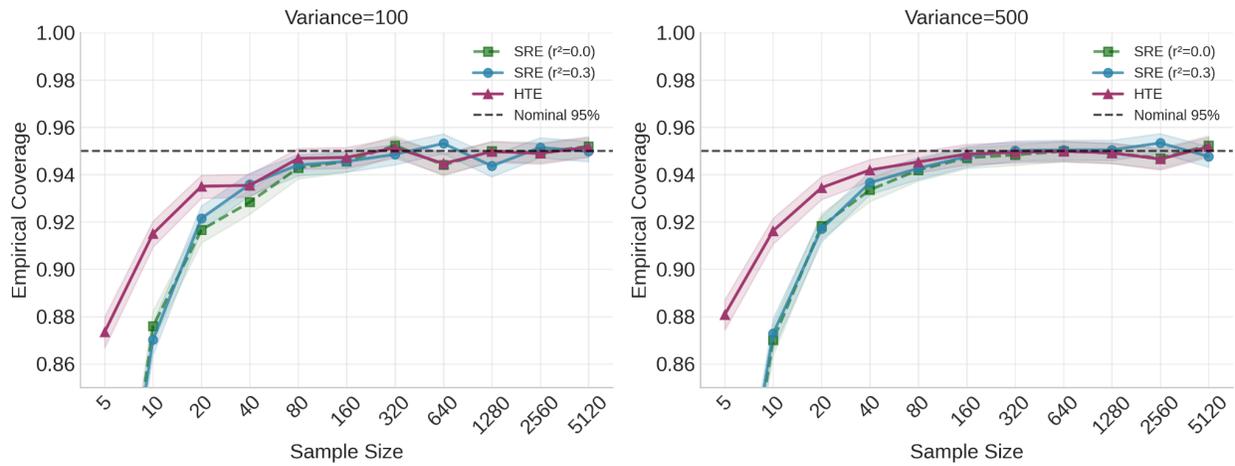

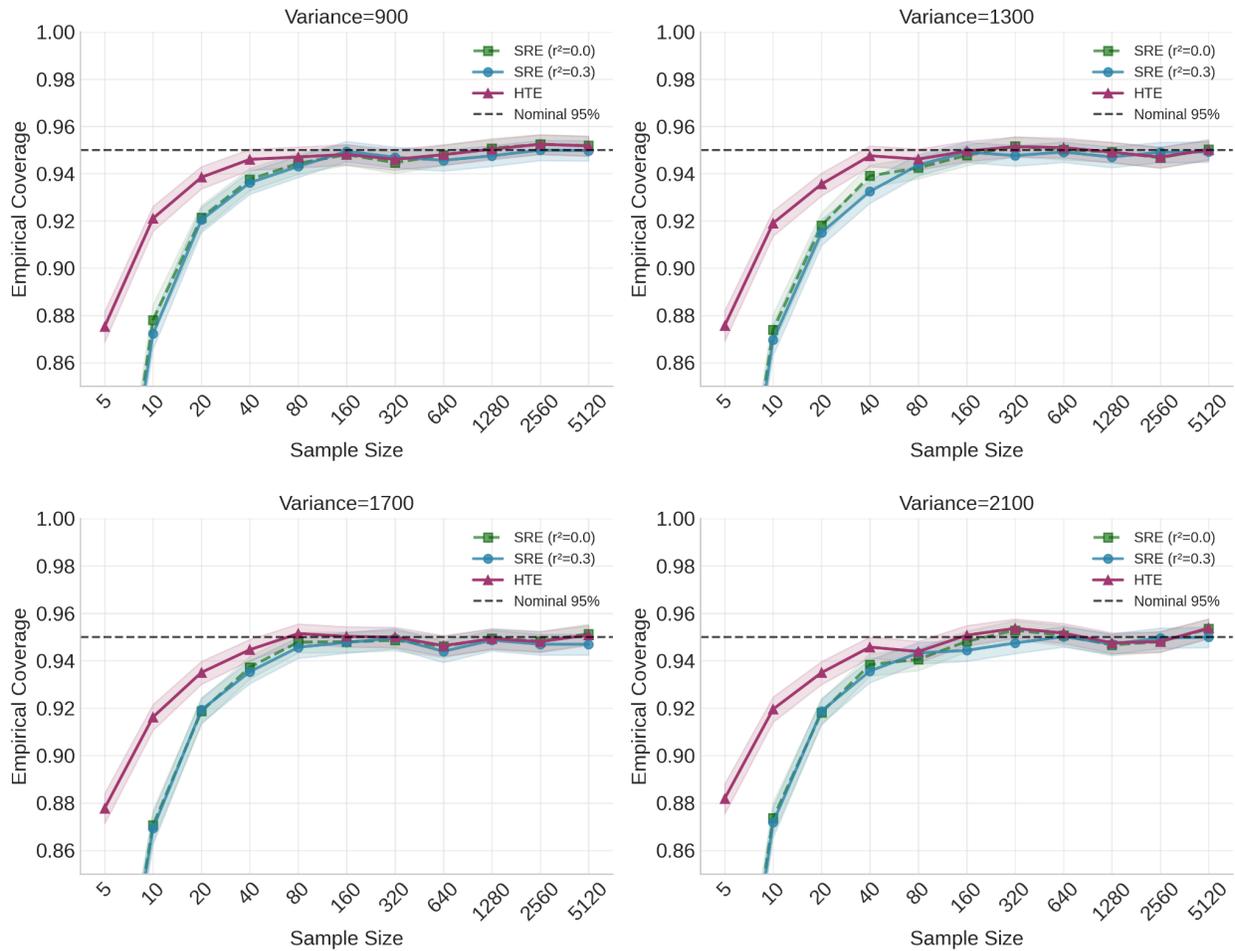

**Figure 4.** Empirical coverage probabilities for all estimators (SRE-uncorr, SRE-corr, and HTE) along with their MC uncertainties (y-axis) under various sample sizes (x-axis) and population variances ∈ {100, 500, 900, 1300, 1700, 2100} tC²/ha². The horizontal baseline indicates the nominal coverage probability of 95%.

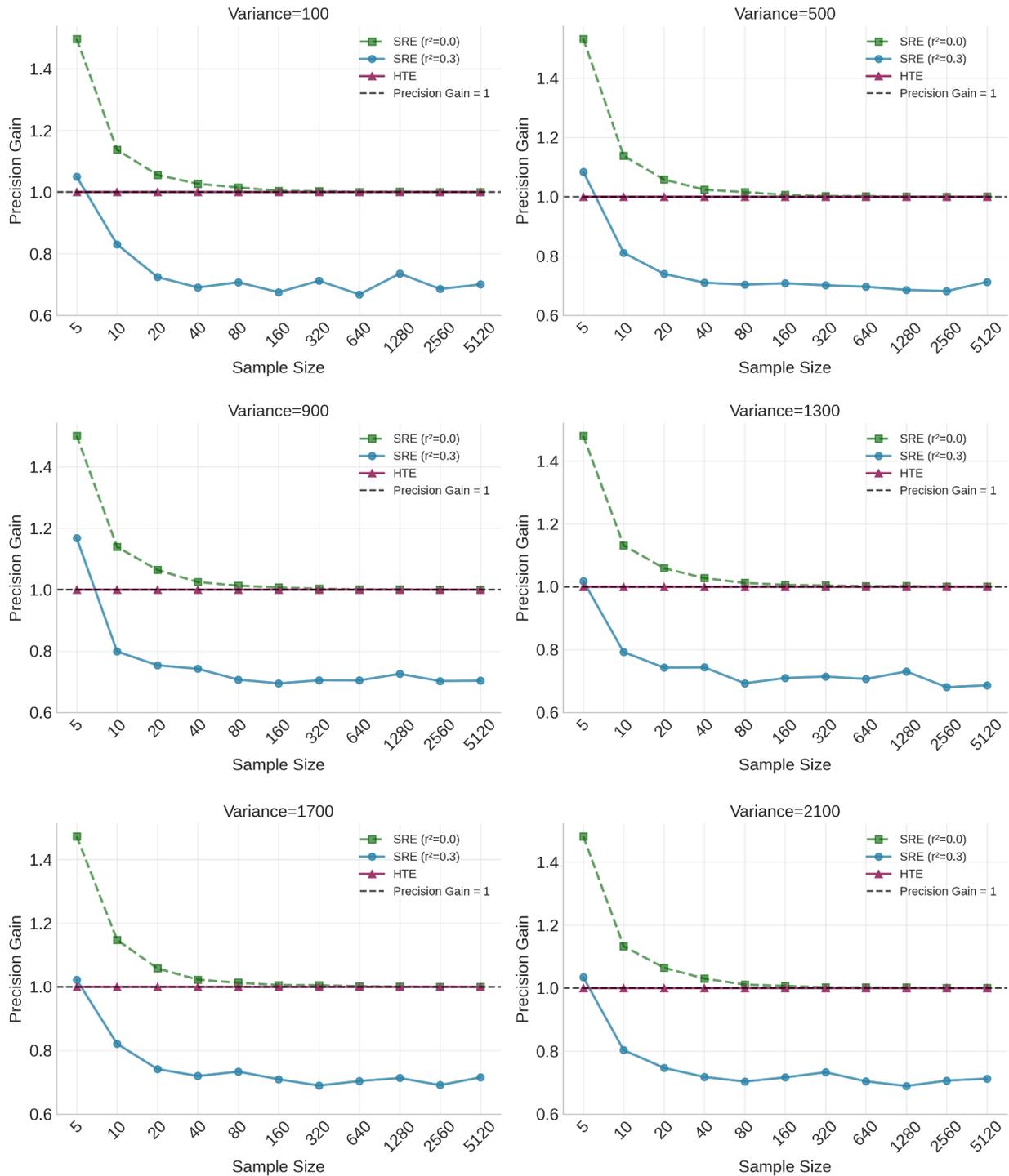

**Figure 5.** Precision gain (y-axis) under various sample sizes (x-axis) and population variances ∈ {100, 500, 900, 1300, 1700, 2100} tC²/ha². Comparison is made across all simulated estimators, i.e., SRE-uncorr, SRE-corr, and the HTE.